\documentclass[aps,groupaddress,pra,twocolumn,showpacs,letterpaper,tighten,float,longbibliography,nofootinbib]{revtex4-1}
\usepackage{amssymb}
\usepackage{amsbsy}
\usepackage{amsmath}
\usepackage{epsfig}
\usepackage{graphicx}
\usepackage{array}
\usepackage[utf8]{inputenc}
\usepackage{textcomp}
\usepackage{color}
\usepackage{braket}
\usepackage{bm,hyperref}
\usepackage[titletoc,toc,title]{appendix}

\definecolor{myblue}{rgb}{.93, .93, 1}

\newcommand{\beq}{\begin{equation}}
\newcommand{\eeq}{\end{equation}}
\newcommand{\p}{\partial}

\begin{document}

\title{Emergent Phases of Fractonic Matter}
\author{Abhinav Prem}
\email{abhinav.prem@colorado.edu}
\author{Michael Pretko}
\author{Rahul M. Nandkishore}

\affiliation{Department of Physics and Center for Theory of Quantum Matter, University of Colorado, Boulder, Colorado 80309, USA}

\date{\today}

\begin{abstract}
Fractons are emergent particles which are immobile in isolation, but which can move together in dipolar pairs or other small clusters. These exotic excitations naturally occur in certain quantum phases of matter described by tensor gauge theories.  Previous research has focused on the properties of small numbers of fractons and their interactions, effectively mapping out the ``Standard Model" of fractons.  In the present work, however, we consider systems with a finite density of either fractons or their dipolar bound states, with a focus on the $U(1)$ fracton models.  We study some of the phases in which emergent fractonic matter can exist, thereby initiating the study of the ``condensed matter" of fractons.  We begin by considering a system with a finite density of fractons, which we show can exhibit microemulsion physics, in which fractons form small-scale clusters emulsed in a phase dominated by long-range repulsion.  We then move on to study systems with a finite density of mobile dipoles, which have phases analogous to many conventional condensed matter phases.  We focus on two major examples: Fermi liquids and quantum Hall phases.  A finite density of fermionic dipoles will form a Fermi surface and enter a Fermi liquid phase.  Interestingly, this dipolar Fermi liquid exhibits a finite-temperature phase transition, corresponding to an unbinding transition of fractons.  Finally, we study chiral two-dimensional phases corresponding to dipoles in ``quantum Hall" states of their emergent magnetic field.  We study numerous aspects of these generalized quantum Hall systems, such as their edge theories and ground state degeneracies.
\end{abstract}

\maketitle



\section{Introduction}
\label{intro}

Quantum phases of matter with long-range entanglement, such as spin liquids and fractional quantum Hall systems, are strikingly characterised by the presence of fractionalized quasiparticles.  As a familiar example, two-dimensional systems can host anyon excitations, characterized by their non-trivial braiding statistics.  In the presence of symmetries, these excitations can carry fractional quantum numbers, exhibiting the phenomenon of symmetry fractionalization. Perhaps the most famous manifestation of this behavior occurs in the celebrated $\nu = 1/3$ Laughlin fractional quantum Hall state~\cite{Laughlin}, where the charge $e/3$ quasi-particles have been directly observed in experiment~\cite{Kane1994,Picciotto,Saminadayar}. The phenomenon of fractionalization has been studied in great detail, specifically in the context of symmetry enriched topological (SET) phases~\cite{wen2002,levin,essin,ran,lu,barkeshli,xu,hung}.  It is now well-established that the appropriate theoretical framework for understanding fractionalization is that of gauge theories, which also describe the fundamental forces of our universe. The resulting theoretical developments on fractionalization have led to a fruitful exchange of ideas between the condensed matter and high energy communities.

The most familiar types of gauge theories can be formulated in terms of a vector gauge field $\vec{A}$, just as in ordinary electromagnetism. Such vector gauge fields account not only for all gauge-mediated interactions in the Standard Model, but also for the theory of the fractional quantum Hall effect~\cite{wenbook}, superconductors with dynamical electromagnetism~\cite{sondhi,moroz}, and most known examples of spin liquid states~\cite{wen2002}.  As such, treatments of fractionalization have historically focused almost exclusively on vector gauge fields.  However, there is no reason in principle why the gauge field must transform as a vector object under rotations.  Motivated by this, recent work has set out to study fractionalization patterns described by more general tensor gauge fields~\cite{rasmussen,sub,genem}.  For instance, a quantum phase of matter could be described by a two-index tensor $A_{ij}$, or a tensor of even higher rank.

Gauge theories with such a tensor gauge field can describe a radically different form of fractionalization from that occurring in conventional vector gauge theories.  Whereas vector gauge theories only feature fractionalization of internal quantum numbers, like charge or spin, particles coupled to a tensor gauge field can exhibit fractionalization of the ability to move through space.  The most notable example of this phenomenon is the existence of ``fracton" excitations in certain tensor gauge theories.  These new particles, first seen in the context of exactly solvable spin models~\cite{chamon,bravyi,castelnovo,haah,haah2,yoshida,fracton1,fracton2}, have no ability to move by themselves, i.e. an isolated fracton is strictly immobile.  Nevertheless, when a fracton combines with an appropriate number of other fractons, it can form a mobile bound state which is free to move around the system~\footnote{The smallest mobile bound state of fractons is usually also a non-trivial excitation of the system, which cannot decay directly into the vacuum, and therefore exists as a stable particle.  The exception to this is in certain ``fractal" fracton models, such as Haah's code, where all mobile bound states are trivial~\cite{haah}.}. In this sense, a fracton is only a fraction of a conventional mobile excitation.  The condensed matter literature has seen a flurry of recent activity fleshing out the properties of these strange new particles~\cite{williamson,han,sagar,prem,mach,hsieh,slagle1,decipher,screening,nonabelian,balents,chiral,slagle2,albert,devakul,regnault,regnault2,albert2,leomichael}. 

More generally, the restriction on mobility of charges in tensor gauge theories need not occur in all directions.  There are other types of ``subdimensional particles" which are immobile in only certain directions~\cite{sub}.  For example, some tensor gauge theories host one-dimensional particles, which are free to move only along a particular line.  The physics of tensor gauge theories is therefore much richer than simply the theory of fractons.  Nevertheless, fracton excitations have been studied in more detail than the other members of the subdimensional family, and we will focus in this work on fractons and their bound states.

Essentially all research in this area to date has focused on the properties of fractons in isolation, or in the presence of a small number of other fractons.  Such treatments are, in a sense, determining the fundamental particle physics governing the behavior of fractons.  For example, previous work has formulated the generalized electromagnetic interactions between fractons and $U(1)$ tensor gauge fields~\cite{genem}.  We can regard all of these previous studies as mapping out the ``Standard Model" of fractons.

While the fundamental behavior of fractons has been studied in detail, there has been comparatively little work studying the behavior of a system with a finite density of fractons.  This is an important problem to study, since a system with emergent tensor gauge structure will not necessarily be at zero chemical potential of fractons.  In fact, there are multiple different chemical potentials to consider.  As we will review, these tensor gauge theories have higher moment charge conservation laws, beyond the conventional monopolar charge conservation law.  As a simple example, some systems exhibit conservation of both charge and dipole moment of fractons, which leads directly to their immobility.  This extra conservation law allows us to consider a second type of chemical potential in the system.  Even when there is zero net charge density of fractons, we can still tune a dipolar chemical potential to obtain a finite density of dipoles.  To fully understand the behavior of fracton systems, we must therefore study not only finite densities of fractons, but also finite densities of the non-trivial bound states.  In doing so, we will be formulating the ``condensed matter theory" of fractons, mapping out the phases in which emergent fractonic matter can exist.  In this paper, we will content ourselves with studying fracton systems which have two non-trivial types of particles: immobile fractons and fully mobile dipoles.  Many of these principles can be carried over to future studies of more complicated systems.

We will begin by studying the case of a finite density of fractons.  This may seem trivial at first, since fractons tend to be locked in place.  However, while a fracton in isolation is strictly immobile, multiple fractons are capable of limited motion via ``pushing off" of each other, leading to a mutual sense of inertia, in a manifestation of Mach's principle~\cite{mach}.  In particular, for a system of fractons at density $\rho$, the fractons will possess a finite effective mass, $m\sim \rho^{-1}$.  While the fractons have now lost their characteristic immobility, there is still one crucial difference between fractons and conventional mobile particles.  Like-charged fractons in a $U(1)$ tensor gauge theory experience two types of forces: a long-range repulsion mediated by the gauge field~\cite{genem}, and an effective ``gravitational" attraction which is generically short-ranged~\cite{mach}.  This type of situation, with short-range attraction and long-range repulsion, provides the natural conditions for microemulsion physics.  The fractons will bind into small-scale clusters dominated by the short-range attraction, which in turn act as droplets ``emulsed" in a phase dominated by the long-range repulsion.  At low densities, the system will form a Wigner crystal of such fracton clusters.  As the strength of the repulsion is increased, the size of clusters will decrease until the system becomes a Wigner crystal of individual fractons.  We will determine the necessary conditions on the repulsive potential for microemulsion physics to hold and will estimate the typical size of fracton clusters.

We will then move on to the study of systems with zero fracton density, but with a finite density of dipoles.  We will assume throughout that dipole moment is quantized (such that there is a minimal dipole moment) as is always the case when the theory arises from an underlying lattice model.\cite{sub}  Dipoles are intrinsically mobile particles, which allows their phases of matter to be studied through more conventional means.  Furthermore, these dipoles can have either bosonic or fermionic statistics.  We can therefore imagine putting these mobile dipoles into almost any phase encountered in conventional condensed matter.  In this paper, we will focus on the dipolar analogues of two familiar phases of matter: Fermi liquids in three-dimensions (3D) and quantum Hall phases in two-dimensions (2D).

We study a system with a finite density of fermionic dipoles, primarily focusing on the case of a single species of dipole ($i.e.$ all polarized in one direction) in 3D.  In this case, we expect the dipoles to form a Fermi surface.  In the fracton model we focus on, the dipoles interact with a $1/r$ repulsive interaction, which will be screened at finite density, just like the conventional Coulomb interaction between electrons.  All of the usual arguments for Fermi liquid theory can be carried over esssentially unchanged.  One notable feature of this dipolar Fermi liquid is the behavior of fractons.  We show that fractons in such a system have a logarithmic interaction energy.  This leads to a finite-temperature phase transition, corresponding to the unbinding of fractons, in similar spirit to a BKT transition.  In the low-temperature phase, the presence of a sharp Fermi surface will result in Friedel oscillations in the spin density, which is a useful experimental diagnostic.  When this phase is realized in a weak Mott insulator ($i.e.$ close to the metal-insulator transition), one should also be able to observe Friedel oscillations in the charge density.

Finally, we will consider systems in two spatial dimensions which have both a finite density of dipoles and a non-zero expectation value of the emergent magnetic field associated with the tensor gauge field.  In this case, the mobile dipoles will respond to this field much like an electron would respond to an external magnetic field, forming the dipolar analogue of a quantum Hall state~\footnote{Importantly, there is no quantum Hall effect of normal electrical conductivity.  Physically, these states will be most likely to occur in Mott insulating spin liquids.}.  We will show that these dipolar quantum Hall states fit naturally into the framework of the recently discovered chiral fracton phases, described by tensor Chern-Simons theories~\cite{chiral}.  We will study many of the natural questions associated with these generalized quantum Hall states, such as their level quantization and ground state degeneracy.  We will also demonstrate the existence of gapless chiral edge modes, which will result in a robust thermal Hall effect.


\section{Review of $U(1)$ Fractons}
\label{review}

Recently, in a series of papers~\cite{sub,genem,mach,screening,chiral}, one of the authors (M.P.) has worked out the properties of 3D U(1) tensor gauge theories, which host fractons and other subdimensional excitations.  These theories provide the natural analogue of the discrete fracton theories formulated by Vijay, Haah, and Fu~\cite{fracton1,fracton2}.  We will here review the simplest example of a $U(1)$ tensor gauge theory---the ``scalar charge theory"---in order to illustrate the main principles underlying these phases.

Instead of a conventional vector gauge field $A^i$, this theory is formulated in terms of a rank 2 symmetric tensor gauge field $A^{ij}$ and its canonical conjugate, $E^{ij}$.  The properties of the theory are almost entirely determined by the form of the Gauss's law for the theory, which takes the form~\footnote{In this paper, we use a notation where Greek indices vary over space-time components ($\mu=0,1,2,3$) and Latin indices are used for spatial components only ($i=1,2,3$). Repeated indices will be implicitly summed over and we will work in units where $e = \hbar = c = 1$, with $c$ being the velocity of the gauge mode.}
\beq
\p_i \p_j E^{ij} = \rho.
\eeq
Whereas the conventional Gauss's law only leads to the conservation of charge, this new Gauss's law has two associated conservation laws,
\beq
\int \rho = \text{const.},\quad \int \vec{x}\rho = \text{const.},
\eeq
corresponding to conservation of charge and dipole moment respectively.  This extra conservation law has a severe consequence for the charges of the theory.  A single charge cannot move while conserving the dipole moment of the system.  Therefore, isolated charges in this system are locked in place and are fracton excitations.  A charge can only move if it combines with an opposite charge to form a dipolar bound state, which is free to move around the system.  Such dipoles are themselves non-trivial objects, since dipole conservation prevents them from decaying directly into the vacuum.

Within the low-energy sector, where $\partial_i\partial_j E^{ij} = 0$, the system is invariant under the following gauge transformation
\begin{equation}
A_{ij}\rightarrow A_{ij} + \partial_i\partial_j \alpha,
\end{equation}
for gauge parameter $\alpha(\vec{x})$ with arbitrary spatial dependence.  The most relevant ``magnetic field" object consistent with this gauge transformation takes the form of a non-symmetric traceless rank 2 tensor
\beq
B_{ij} = \epsilon_{iab}\,\p^a A^b_{\,\,\,j} .
\eeq
In terms of the electric and magnetic fields, the Hamiltonian for this theory is given by
\beq
\mathcal{H} = \int \left(\frac{1}{2} E^{ij}E_{ij} + \frac{1}{2}B^{ij}B_{ij} + A^{ij}J_{ij} \right),
\eeq
where $J_{ij}$ is a symmetric current tensor describing the motion of fractons.  Note that, whereas normal particles have a vector current describing their motion, fractons can only move through multi-body hopping processes, which are conveniently captured by a symmetric tensor satisfying
\beq
\p_t \rho + \p_i\p_j J^{ij} = 0.
\eeq
which serves as the generalized continuity equation of the theory~\cite{genem}.

This theory can also be formulated in Lagrangian language as
\beq
\mathcal{L} = \frac{1}{2}\left(\dot{A}_{ij} - \p_i\p_j\phi \right)^2 - \frac{1}{2}B^{ij}B_{ij} - A^{ij} J_{ij} - \phi \rho,
\eeq
where dots denote temporal derivatives and where $\phi$ is a field analogous to the temporal component, $A_0$, of the more familiar rank 1 gauge theory, serving as a Lagrange multiplier enforcing Gauss's law~\cite{chiral}.  (Note that this theory does not have Lorentz invariance, so $\phi$ does not transform as a ``0 component" of the gauge field.)  In this language, we can write a more general time-dependent gauge transformation within the low-energy sector
\beq
A_{ij}\to A_{ij} + \p_i\p_j \alpha,\quad \phi \to \phi + \dot{\alpha}
\eeq
for scalar field $\alpha(\vec{x},t)$ with arbitrary dependence on space and time. 

One curious feature of this particular model is the fact that the interfracton potential grows linearly, $V(r)\sim r$.  (This is not generic to $U(1)$ fractons, which in other models have a standard decaying potential.)  In conventional vector gauge theories, a linear interparticle potential is indicative of an instability to a gapped confined phase.  In this theory, however, there is a stable gapless phase, regardless of the large energy cost necessary to separate particles~\cite{rasmussen,sub}.  Furthermore, once this energy cost has been paid, the immobility of fractons stabilizes them from collapsing directly back into the vacuum.  Nevertheless, the linear energy cost indicates that these fractons cannot be thermally excited in large numbers.  We will see later, however, that the presence of a dipolar Fermi surface can screen the linear potential down to a logarithmic interaction, which will allow for the proliferation of fractons above a certain temperature.

\section{Microemulsions of Fractons}

We begin by considering a three-dimensional system which has a finite density of $U(1)$ fractons.  At finite density, fractons endow each other with inertia through the virtual exchange of dipole moment, and their characteristic immobility disappears~\cite{mach}.  Even though the fractons can now freely move around the system, there is still one crucial feature which sets fractons apart from conventional mobile particles.  A fracton will lower its effective inertia, and thereby move more quickly, when it is in the immediate vicinity of another fracton.  The result is an always-attractive geometric force between fractons, which plays the role of an effective gravitational interaction.  As shown in previous work~\cite{mach}, the gravitational attraction between fractons is generically short ranged in the models studied in the condensed matter literature.  The effective short-ranged attractive potential takes the form
\begin{equation}
V_s(r) = -V_0 e^{-Mr},
\end{equation}
where $M$ is the mass scale of the mobile dipoles and $V_0$ is a constant.

While the emergent gravitational force provides a short-range attraction, this is not the only interaction between fractons.  Like-charged fractons in $U(1)$ models also exhibit a conventional gauge-mediated long-range repulsion.  The precise power law of this repulsive interaction depends on the model, but we can readily identify an interesting universal feature which holds for a range of different potentials.  A model with short-range attraction and long-range repulsion provides precisely the sort of conditions necessary for microemulsion physics (for discussions of microemulsion physics in a more traditional context, see \cite{JameiSpivakKivelson, SpivakKivelson, ParameswaranMicroemulsion}).  At short distances, fractons attract each other and will have a tendency to bind together into clusters.  At longer distances, however, we expect the power-law repulsion to take over, preventing the fractons from coalescing into a single large cluster.  Instead, fracton clusters will have some typical intermediate size and will behave as mesoscopic ``particles" with an effective repulsive interaction.  This repulsion will keep the fracton clusters emulsed in the surrounding medium, instead of phase separating into a single large cluster.  The situation is reminiscent of protons in a nucleus, held together by short-range attraction, which interact with other nuclei through a long-range Coulomb repulsion.  At low densities, the system will form a Wigner crystal of fracton clusters.

This physical picture, while appealing, will turn out to hold only for a certain range of potentials.  If the repulsion is too weak, all fractons will collapse into a single cluster.  If the repulsion is too strong, all clusters will break apart into a Wigner crystal of individual fractons.  In order to make more concrete statements, we must consider the precise form of the repulsive interaction.  We will break up the analysis into two classes of repulsive power-law potentials, both of which are relevant in fracton phases.

\subsection{Decaying Potentials}

\begin{figure}[t!]
 \centering
 \includegraphics[width=0.45\textwidth]{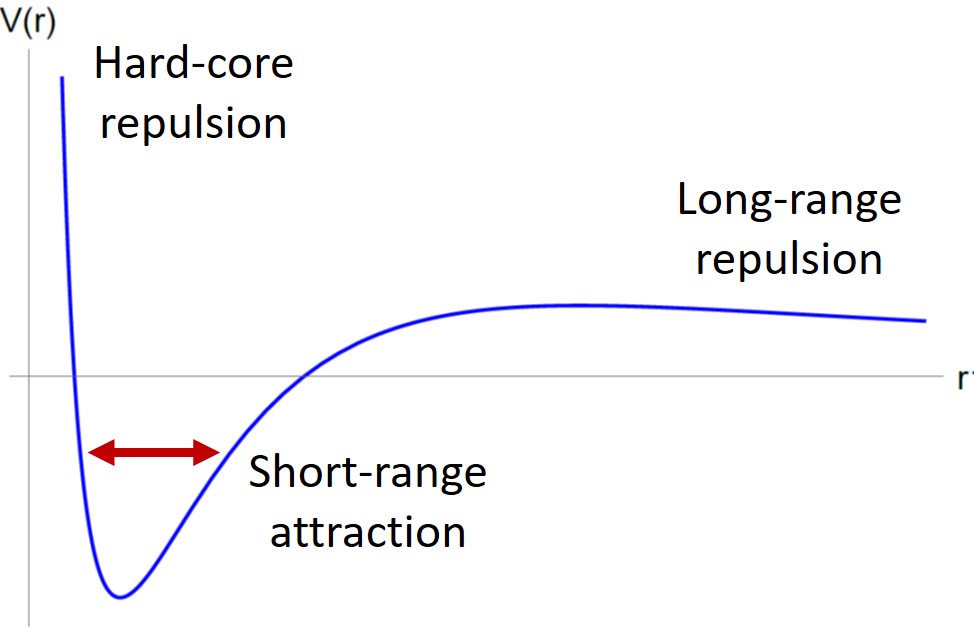}
 \caption{$V(r)$ vs $r$.  The decaying potential has both long-range and ``hard-core" repulsions, with a region of attraction just outside the core.}
 \label{fig:potential}
 \end{figure}

We first assume that, in addition to the short range attraction, the fractons have a conventional decaying repulsive potential
\begin{equation}
V_l(r) = \frac{\alpha}{r^n},
\end{equation}
for some power $n$.  This is the situation which holds in some fracton models, such as the Gu-Wen emergent gravity model, for which $n=1$ ($i.e.$ a Coulomb potential)~\cite{gu1,gu2}.  Such a potential provides both a long-range repulsion and also a ``hard-core" repulsion at the shortest distances, with the short-range interaction providing an attraction immediately outside the core (see Figure~\ref{fig:potential}).

\begin{figure}[t!]
 \centering
 \includegraphics[scale=0.4]{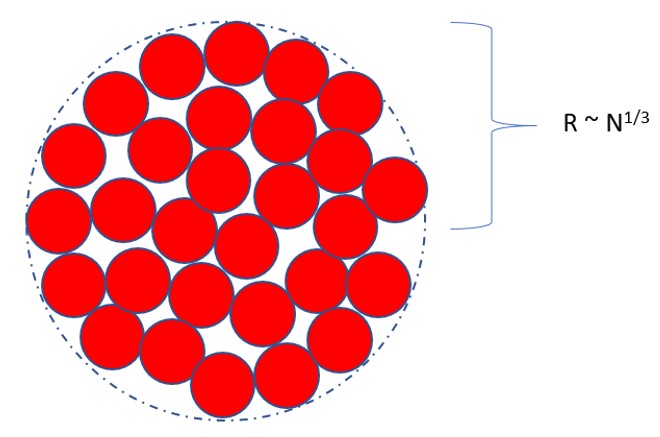}
 \caption{We consider a cluster of fractons that is approximately close-packed, assuming a hard-core radius $a$ for the fractons.  The radius of the cluster scales as $R\sim N^{1/3}$.}
 \label{fig:closepack}
 \end{figure}

On short scales, we expect to see fractons clustering together into bound states.  Taking the fractons to have a hard core of radius $a$, the lowest energy configuration will be approximately close-packed out to some radius $R$, as in Figure~\ref{fig:closepack}.  The total number $N$ of particles in the cluster scales as $N\sim(R/a)^3$.  In order to determine the most energetically favorable value for $N$ (and thereby $R$), we need to estimate the contributions to the energy of the cluster from both the short-range and long-range potentials.

For the short-range potential, it is sufficient to consider interaction energy between nearest neighbors.  We therefore approximate the potential by $V_s = -V_0$ for nearest neighbor pairs, $V_s=0$ otherwise.  The total contribution to the energy of the cluster from the short-range interaction is then given by
\begin{equation}
E_s = -c_1V_0 N + c_2V_0N^{2/3},
\end{equation}
where $c_1$ and $c_2$ are positive numbers of order unity.  The first term represents the interaction energy of particles in the bulk, with $c_1$ quantifying the number of nearest neighbors.  The positive $N^{2/3}$ term represents the number of particles on the surface, which do not get all the energetic benefits of particles in the bulk, due to having fewer nearest neighbors.  Keeping this surface term will turn out to be crucial.

In addition to the short-range interaction energy, we also need the contribution to the energy from the long-range repulsive potential.  This interaction energy behaves as $E_l \sim \alpha N^2/R^n \sim \alpha N^{2-\frac{n}{3}}/a^n$, so we write
\begin{equation}
E_l = c_3\frac{\alpha}{a^n}N^{2-\frac{n}{3}},
\end{equation}
for $c_3$ of order unity.  The total energy of the cluster is then given by
\begin{equation}
E = -c_1V_0N + c_2V_0N^{2/3} + c_3\frac{\alpha}{a^n}N^{2-\frac{n}{3}},
\end{equation}
and the energy per fracton is
\begin{equation}
\frac{E}{N} = -c_1V_0 + c_2V_0N^{-1/3} + c_3\frac{\alpha}{a^n}N^{1-\frac{n}{3}}.
\end{equation}
We can find the most energetically favorable configuration of the whole system by minimizing the energy per fracton
\begin{equation}
\frac{d(E/N)}{dN} = -\frac{1}{3}c_2V_0N^{-4/3} + c_3(1-\frac{n}{3})\frac{\alpha}{a^n}N^{-n/3}
\end{equation}
\begin{equation}
\Rightarrow N_c \sim \bigg(\frac{V_0a^n}{\alpha(1-\frac{n}{3})}\bigg)^{\frac{3}{4-n}}\sim \bigg(\frac{V_s(a)}{V_l(a)(1-\frac{n}{3})}\bigg)^{\frac{3}{4-n}}.
\end{equation}
In other words, the optimal particle number of the cluster is determined by the ratio of the short-range and long-range interaction energies of neighboring fractons.  When the total number of fractons in the system is less than $N_c$, all of the fractons will clump together into a single cluster.  At larger particle numbers, the system will have multiple clusters, interacting with each other through the long-range repulsion.  The typical size of these clusters is given by
\begin{equation}
R \sim aN_c^{1/3} \sim a\bigg(\frac{V_s(a)}{V_l(a)(1-\frac{n}{3})}\bigg)^{\frac{1}{4-n}}.
\end{equation}
It is worth noting that, for $n\geq 3$, the cluster size blows up.  This indicates that a long-range repulsion weaker than $1/r^3$ is no longer enough to keep the fractons from phase-separating into a single cluster, in a form of ``gravitational collapse" of the system~\footnote{Note that the hard core prevents the fractons from further collapsing into a black hole.  Whether or not black hole physics is accessible within the fracton framework remains an important open question.}. We therefore see that only repulsive interactions with $n<3$ will exhibit microemulsion physics.

\begin{figure}[t!]
 \centering
 \includegraphics[width=0.5\textwidth]{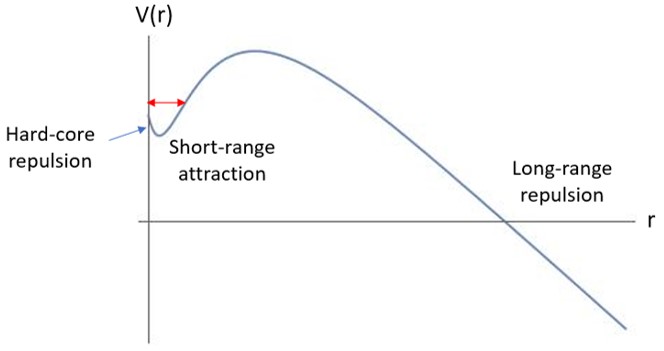}
 \caption{$V(r)$ vs $r$.  The growing potential has a similar profile to the decaying case, except that the long-range repulsive potential now grows unbounded in magnitude, destabilizing the microemulsion.}
 \label{fig:potential3}
 \end{figure}

\subsection{Growing Potentials}

While some $U(1)$ fracton models feature a conventional decaying potential, other models have a repulsive potential which increases in magnitude as the fractons are separated.  For example, we have already discussed that the scalar charge theory exhibits a linear repulsive potential between fractons, $V = -\alpha r$.  More generally, we can consider a growing repulsive potential of the form
\begin{equation}
V_l(r) = -\alpha r^n.
\end{equation}
When combined with short-range effects, this produces the potential energy profile seen in Figure \ref{fig:potential3}.  For a potential of this form, the system can always lower its energy by breaking apart a cluster into a configuration with well-separated fractons.  This becomes readily apparent if we attempt to determine the typical cluster size using the strategy of the previous section.  All previous statements about the short-range interaction carry over directly.  The long-range interaction is slightly trickier.  This contribution to the energy behaves as
\begin{equation}
E_l \sim -\alpha N^2 (R^n - L^n),
\end{equation}
where $L$ is the system size.  We are here writing the energy of the cluster relative to a state in which the fractons are well-separated, which has energy of order $-\alpha N^2 L^n$.  With respect to this reference point, the total energy of a cluster is given by
\begin{equation}
E = -c_1V_0 N + c_2V_0 N^{2/3} + c_3 \alpha L^n N^2 - c_4 \alpha a^n N^{2+\frac{n}{3}}.
\end{equation}
The energy per particle is given by
\begin{equation}
\frac{E}{N} = -c_1V_0 + c_2V_0N^{-1/3} + c_3\alpha L^n N - c_4 \alpha a^n N^{1+\frac{n}{3}}.
\end{equation}
Since $L\gg a$, the fourth term is negligible compared to the third, so we can drop it.  (The fourth term only becomes relevant when $N\sim (L/a)^3$, at which point the notion of separate clusters breaks down anyway.)  Within this approximation, we can find the optimal cluster size by minimizing the remaining terms with respect to $N$,
\begin{equation}
\frac{d(E/N)}{dN} = -\frac{1}{3}c_2V_0N^{-4/3} + c_3\alpha L^n
\end{equation}
\begin{equation}
\Rightarrow N_c \sim \bigg(\frac{V_0}{\alpha L^n}\bigg)^{3/4}.
\end{equation}
If we take the thermodynamic limit, $L\rightarrow\infty$, we see that the typical cluster size vanishes, $N_c\rightarrow 0$, indicating that the repulsion has caused all clusters to break apart.  The resulting state will feature fractons which are spaced apart as much as possible, in a Wigner crystal configuration.

We have now seen that a repulsive potential $V\sim r^n$ for $n>0$ leads to no clusters at all in the system, whereas a potential $V\sim r^{-n}$ for $n \geq 3$ leads to the formation of a single phase-separated cluster.  We can therefore conclude that microemulsion physics, with finite size clusters emulsed in a Wigner crystal of clusters, only holds for repulsive potentials $V\sim r^{-n}$ with $0<n<3$.  Stronger repulsions will result in a single-particle Wigner crystal, while weaker repulsions will result in the ``gravitational collapse" of the fractons.

\section{Dipolar Fermi Liquids}

We now move on to study a system which does not have a finite density of fractons, but rather a finite density of mobile dipoles.  We will focus our attention specifically on the dipoles of the scalar charge theory, discussed in Section~\ref{review}.  We will assume throughout that there is a certain minimal size for dipole moments ($i.e.$ dipole moment is quantized), as is always the case when the scalar charge theory arises from an underlying lattice model~\cite{sub}.  These dipoles can be either bosons or fermions, as discussed in Appendix B.  In this paper, we will focus on fermionic dipoles, which have phases of matter directly analogous to conventional electronic phases.  For a system with a finite density of fermionic dipoles, the simplest fate for the system would seem to be that the dipoles form a Fermi surface, which is the first possibility that we will explore.  We will study a three-dimensional system, where there are no instanton effects which could destabilize the system.  It is possible that a two-dimensional dipolar Fermi surface is stable as well, similar to the conventional 2D spinon Fermi surface~\cite{sungsik}, but this would require a more detailed analysis which we leave to future studies.

We first study the case where there is only a finite density of one species of dipole $p^i$, with a specific orientation of dipole moment.  Recall that dipole moment is a conserved quantity, so an isolated dipole cannot change its orientation.  Furthermore, dipole moment is quantized in the system, so scattering between dipoles cannot change the orientation without paying a large finite energy cost.  We assume that all interaction energies in the problem are small compared to this scale.  It is then valid to consider a system of only $p^i$-oriented dipoles.  Of course, such a polarized state breaks any rotational or inversion symmetries of the system.  In terms of the microscopic degrees of freedom from which these dipoles emerge ($e.g.$ spins in a spin liquid), a single-species dipolar Fermi liquid will be a state in which symmetry breaking and long-range entanglement coexist.

\subsection{Justification of Fermi Liquid Theory}

A finite density of noninteracting fermions will always form a Fermi surface.  But in order to justify the existence of a stable interacting Fermi liquid, we must examine precisely how these dipoles interact with each other.  We will confine our attention to the scalar charge theory discussed earlier.  From the generalized electromagnetism of this model~\cite{genem}, we know that the interparticle potential between two dipoles, $p$ and $p'$, takes the form
\begin{equation}
V(r) = \frac{(p\cdot p')}{8\pi r} - \frac{(p\cdot r)(p'\cdot r)}{8\pi r^3}.
\label{int}
\end{equation}
For identical dipoles, $p = p'$, this reduces to
\begin{equation}
V(r) = \frac{p^2\sin^2\theta}{8\pi r},
\label{interact}
\end{equation}
where $\theta$ is the angle between $p$ and $r$.  The corresponding force between identical dipoles is generically repulsive, except for a line of zero force at $\theta = 0$.  Importantly, the force is never attractive.  Also, we note that the $1/r$ potential (and corresponding $1/r^2$ force) scales exactly like the normal Coulomb interaction between electrons.  We can hence essentially regard the interaction between dipoles as simply an anisotropic Coulomb force.  Then, just as in normal Fermi liquid theory, the dipoles will be able to screen each other.  (The details of dipolar screening are worked out explicitly in Appendix A.)  After accounting for screening, the resulting screened quasiparticles will only have weak short-range interactions.

At this point, the dipole moment of the fermions becomes mostly irrelevant to the problem.  We have a system of fermions with short-range interactions, with the dipole moment simply serving as an extra internal quantum number which has no effect on the traditional Fermi liquid analysis.  All of the usual interesting aspects of Fermi liquid theory will carry over unchanged.  There will be a discontinuity in dipole occupation number at a sharp Fermi surface in momentum space (with a quasi-particle residue $Z<1$).  Also, for appropriate values of Landau parameters, the system will host a zero sound mode, representing oscillations of the Fermi surface, which provides a way to distinguish the system from a free Fermi gas of dipoles.

\begin{figure}[t!]
 \centering
 \includegraphics[scale=0.4]{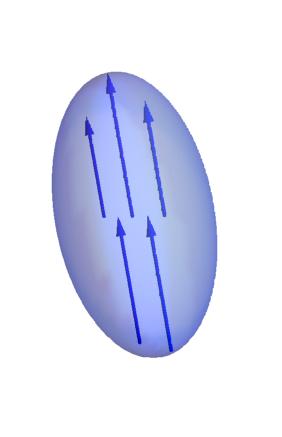}
 \caption{The anisotropic nature of the interaction between dipoles will cause the Fermi surface to elongate in the direction of the dipole moment, forming roughly a prolate spheroid.}
 \label{fig:prolate}
 \end{figure}

While the dipole quantum number does not significantly affect the Fermi liquid analysis, there is one important way in which it makes its presence known in the low-energy physics.  While the bare interaction of Eq.~\eqref{interact} is screened, it remains highly anisotropic, with a strong repulsion between side-by-side dipoles and zero interaction between end-to-end dipoles.  As such, the dipoles will tend to be arranged more densely in the direction of their dipole moment.  This corresponds to a larger Fermi momentum $k_F$, in this direction than in the two perpendicular directions.  Thus, the anisotropic interaction between dipoles will lead to a Fermi surface which is elongated along the direction of the dipole orientation.  Starting from a nearly isotropic system, the interactions will cause the Fermi surface to roughly take the shape of a prolate spheroid, as illustrated in Figure~\ref{fig:prolate}.

This elongation should manifest itself in the Friedel oscillations of the system, which will have a shorter wavelength in the direction of the dipole moment.  When this dipolar phase of matter is realized in a Mott insulating spin liquid, these Friedel oscillations will be most prominently seen in the spin density, since the dipoles will carry spin but not charge.  However, for a weak Mott insulator (close to the metal-insulator transition), the coupling between the charge and spin sectors is strong enough to observe Friedel oscillations in the charge density as well, as seen in certain Mott transitions with a ``ghost" Fermi surface~\cite{ghost}.

It is worth noting that, as in more familiar $U(1)$ spin liquids with spinon Fermi surfaces, the Fermi velocity of the dipoles will generically be of the same order as the speed of the gapless gauge mode.  As such, the dynamical screening of this system will be more complicated than that of a normal metal, which should manifest itself in the response functions of the system.  We leave the detailed study of dynamical screening as a problem for future study.

\subsection{Finite-Temperature Phase Transition}

Another important aspect of the dipolar Fermi liquid to investigate is the behavior of fractons, which still occur as excitations of the system (even though they are no longer at finite density).  The bare fractons of the scalar charge theory have a linear interfracton potential, $V\sim r$.  This is the phenomenon of ``electrostatic confinement"~\cite{sub}, which usually makes these fractons irrelevant to the low-energy physics.  In the presence of a finite density of dipoles, however, previous work~\cite{screening} has indicated that the interfracton potential is partially screened.  The present case is slightly different from previous work due to the presence of a Fermi surface and also due to having only one orientation of dipole.  We leave the details of the calculation to Appendix A.  Here, we simply quote the result that the screened potential grows only logarithmically,
\begin{equation}
V_{scr}(r)  \sim  \frac{1}{\sqrt{g}}\log r,
\end{equation}
where $g$ is the density of states of dipoles at the Fermi surface.  The fractons now interact through a logarithmically increasing potential, which is a much milder sort of growth than the bare linear potential.

The logarithmic potential still results in a significant energy cost for an isolated fracton, scaling as $\log L$, where $L$ is the system size, much like a vortex in a two-dimensional superfluid.  Just as in a superfluid, we expect that the fractons will only proliferate above a certain temperature.  The free energy associated with an isolated fracton in a system of size $L$ will take the schematic form
\begin{equation}
F \sim \bigg(\frac{1}{\sqrt{g}}-k_B T\bigg)\log L.
\end{equation}
Fractons will therefore only proliferate at temperatures above a certain critical temperature, where the free energy per particle becomes negative
\begin{equation}
T_c \sim \frac{1}{k_B\sqrt{g}}.
\end{equation}
Below this temperature, fractons will mostly exist in small bound states, such as dipoles.  Above the transition temperature, fractons will be able to unbind and behave independently, just like in the BKT transition of vortices in a superfluid.  When this happens, the dipoles will lose their integrity and break apart into separate fractons, destroying the Fermi surface.  Furthermore, in such a finite temperature system, fractons lose their characteristic immobility and can move around the system (albeit very slowly)~\cite{screening}.  At this point, all interesting properties of both fractons and dipoles have been lost, and the system is in a trivial phase.

We note that, for a three-dimensional system, the density of states $g$ increases with the size of the Fermi surface, indicating a decrease in $T_c$ as the Fermi surface gets bigger.  In contrast, the Fermi temperature $T_F \sim E_F$ increases with the size of the Fermi surface.  For a sufficiently large dipolar Fermi surface, the transition will happen at temperatures well below the Fermi temperature, $T_c \ll T_F$, so we do not need to worry about thermal smearing of the Fermi surface.  The Fermi surface should remain fairly sharp up until the critical temperature, where the dipoles are destroyed.

\begin{figure}[t]
 \centering
 \includegraphics[scale=0.45]{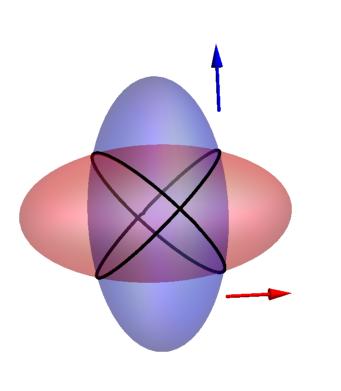}
 \caption{Fermi surfaces of different species of dipoles will be elongated along different directions, which drastically reduces their overlap, shown here by the solid black curves. }
 \label{fig:fermisurface}
 \end{figure}

\subsection{Multi-Species Fermi Liquids}

One might also consider the case where there are finite densities of two or more different orientations of dipoles in the system.  We would like to imagine each species as entering its own Fermi liquid phase, with perhaps some weak coupling between them.  However, the form of the interaction between dipoles (Eq.~\eqref{int}) brings a complication into this picture.  For simplicity, consider the case of two species, $p$ and $p'$, with dipole moments of equal magnitude but in perpendicular directions.  The resulting interparticle potential is
\begin{equation}
V(r) = -\frac{(p\cdot r)(p'\cdot r)}{8\pi r^3} = -\frac{p^2\sin 2\theta}{16\pi r},
\end{equation}
where $\theta$ is the angle between $p$ and the component of $r$ in the $p-p'$ plane.  Note that, whereas the interaction between identical dipoles was strictly repulsive, the interaction between perpendicular dipoles can be either attractive or repulsive depending on the relative positions of the dipoles.  We now have a channel with an attractive interaction, so the system may have pairing instabilities.  A further complication is the fact that the Fermi surfaces of the different dipole orientations will not overlap, as shown in Figure~\ref{fig:fermisurface}, which will make the interspecies pairing problem more intricate.  Ultimately, the study of multi-species dipolar Fermi liquids will be a much more difficult problem than the single-species case, and we leave this to future work.


\section{Dipolar ``Quantum Hall" Phases}
\label{chiral}

The preceding sections of this paper have focused on fracton models in three spatial dimensions, which until recently had been the sole focus of the fracton community, due to folklore that fractons could not occur in less than three dimensions.  The discrete spin models~\cite{fracton1,fracton2} seem to have some fundamental obstruction to being realized in two dimensions (though this issue is not yet fully settled).  Meanwhile, the lattice rotor fracton models~\cite{rasmussen,sub} are described by compact $U(1)$ tensor gauge theories, which are unstable in two dimensions due to instanton effects, just like a conventional compact $U(1)$ gauge field in 2D.

Despite these earlier difficulties, recent work has shown that fractons can exist in two spatial dimensions, and in fact are realized in a simple two-dimensional quantum crystal as disclination defects~\cite{leomichael}.  The elastic theory of such a crystal can be mapped directly onto a \emph{non}compact $U(1)$ tensor gauge theory, which avoids instanton effects, thereby providing an example of a stable fracton model in two dimensions.  This realization opens the door to a whole new class of fractonic phases of matter.  For example, we can place the mobile dipoles into any of the conventional two-dimensional phases of matter.  In this work, we will focus on dipolar analogues of one of the most well-studied types of two-dimensional phases: quantum Hall systems.

Conventional quantum Hall phases can be productively studied through the use of Chern-Simons theories, which capture the flux attachment physics of the composite fermion picture.  Similarly, we should be able to study dipole ``quantum Hall" states by attaching to each dipole some amount of its effective magnetic flux.  Luckily, the appropriate tool for studying such dipolar flux attachment has already been developed, in the form of tensor Chern-Simons theories, first seen in the context of boundary theories to certain three-dimensional fracton models~\cite{chiral}.  We will here apply these generalized Chern-Simons theories to the study of purely two-dimensional dipolar ``quantum Hall" phases.  (We will drop the quotes from here on, with the understanding that these phases do \emph{not} exhibit a quantum Hall response to the physical electromagnetic field.)  By coupling a Maxwell-type tensor gauge theory (arising from elasticity theory, for example) to a tensor Chern-Simons gauge field, we obtain a gapped chiral phase of matter hosting fracton excitations.  We work out some of the basic properties of these new two-dimensional gapped phases, such as their response properties, ground state degeneracies, and edge modes.

We note that these tensor Chern-Simons theories also provide a way to stabilize \emph{compact} $U(1)$ tensor gauge theories in two dimensions.  Such theories can arise, for example, by coupling a two-dimensional quantum crystal to a substrate~\cite{leomichael}.  They can also arise directly in lattice rotor models.  A compact Maxwell tensor gauge theory of this sort has an instability to a trivial gapped confined phase, driven by instanton effects.  However, by adding a tensor Chern-Simons term to the critical point, we can drive the system towards a different nontrivial gapped fixed point, described by dipolar quantum Hall physics.

We emphasize that the $U(1)$ fracton models we consider evade the usual arguments for non-existence of fracton theories in two dimensions.  Such arguments apply to stabilizer code models where the fracton charge is only conserved modulo an integer.  In contrast, we consider only systems of $U(1)$ fractons with an absolutely conserved charge, such as the topological defects in two-dimensional quantum crystals~\cite{leomichael}.  In the following, we will always take the tensor Chern-Simons theories to be coupled to such fractons with conserved charge.

\subsection{Review of Chern-Simons Theory}
\label{csreview}

We begin by reviewing the basics of Chern-Simons field theories described by vector gauge fields, highlighting the elements most relevant for our higher rank generalizations (for a more comprehensive review, we refer the reader to Refs.~\cite{wenbook,tong}.). The structure of these topological quantum field theories (TQFTs) was elucidated in a remarkable paper by Witten~\cite{Witten}, and it has since been realized that they describe the low energy physics of a large class of two-dimensional gapped topological phases of matter, including quantum Hall fluids~\cite{read,wenzee,frohlich}, superconductors~\cite{sondhi,moroz}, and spin liquids~\cite{wen2002}. In particular, they correctly capture the non-trivial ground state degeneracy (GSD) of topological phases on a torus (a direct manifestation of topological order), in addition to the braiding and statistics of fractionalized excitations. While a general discussion should include non-Abelian Chern-Simons theories (which appear e.g., in the context of the $\nu = 5/2$ quantum Hall state~\cite{fradkin} and chiral spin liquids~\cite{wen91}), for simplicity we will restrict our discussion to the Abelian case.

The Chern-Simons action at level $k$ for a vector gauge field $A_\mu$ is 
\beq
\label{CS}
\mathcal{S}_{CS} = \frac{k}{4\pi}\int d^3x\,\epsilon^{\mu\nu\lambda}A_\mu \p_\nu A_\lambda,
\eeq
where $\mu,\nu,\lambda = 0,1,2$. This term is rotationally invariant but breaks parity and time-reversal. Moreover, it is gauge-invariant under $A_\mu \to A_\mu + \p_\mu \alpha$ 
\beq
\mathcal{S}_{CS} \to \mathcal{S}_{CS} + \int d^3x\, \p_\mu \left(\alpha \epsilon^{\mu\nu\lambda}\p_\nu A_\lambda \right),
\eeq
only upto a total derivative term which vanishes on closed manifolds. The Chern-Simons term~\eqref{CS} naturally captures the basic physics of the integer quantum Hall (IQH) effect, since 
\beq
\frac{\delta \mathcal{S}_{CS}}{\delta A_0} = J_0 = \frac{k}{2\pi} B,
\eeq
\beq
\frac{\delta \mathcal{S}_{CS}}{\delta A_i} = J_i = -\frac{k}{2\pi} \epsilon^{ij} E_i.
\eeq
Identifying $k$ with the number of filled Landau levels $\nu$, we see that the Chern-Simons term captures the flux-attachment in the quantum Hall state and describes a Hall conductivity $\sigma_{xy} = \nu/(2\pi)$. While $\nu$ is naturally quantised since it describes the number of filled Landau levels, at first glance $k$ need not be quantized. However, the level $k$ is necessarily quantized as a result of gauge invariance (see Appendix~\ref{quantCS} for details). 

While the above discussion focused only on the integer quantum Hall state, Chern-Simons theory can more generally describe the low-energy physics of phases with fractionalized excitations. The long-distance physics of (Abelian) topological phases is captured by an Abelian Chern-Simons field theory, with the Lagrangian
\beq
\label{EFT}
\mathcal{L} = \frac{1}{4\pi} \epsilon^{\mu\nu\lambda} a^I_{\mu} K_{IJ}\p_\nu a^J_{\lambda}.
\eeq 
Here, $a^I$ is a multiplet ($I=1,2,\dots, N$) of compact U(1) statistical gauge fields and $K_{IJ}$ is a symmetric integer-valued $N\times N$ matrix which encodes the statistics of quasi-particles. The parity of the diagonal entries of the $K$-matrix specify whether the state is fermionic (odd) or bosonic (even). The $K$-matrix describing a gapped $\mathbb{Z}_2$ spin-liquid is given by
\beq
K = \left(
\begin{array}{cc}
0 & 2 \\
2 & 0
\end{array}
\right),
\eeq
while fractional quantum Hall (FQH) states are described by $N = 1$ and $K = m$, with $m$ odd (even) for fermionic (bosonic) states.

As an additional simplification, let us consider this latter case, where $K$ is simply an integer and the theory is governed by a single statistical gauge field $a$,
\beq
\label{sEFT}
\mathcal{L} = \frac{m}{4\pi} \epsilon^{\mu\nu\lambda} a_{\mu} \p_\nu a_\lambda.
\eeq
In order to completely specify the phase, we would also need to specify the quantum numbers carried by quasi-particles, which in the context of FQH states refers to their charge. While this is easily accomplished within the Chern-Simons formalism by adding a mixed Chern-Simons term
\beq
\label{regcoup}
\mathcal{L}_{\text{mixed}} = -\frac{1}{2\pi} \epsilon^{\mu\nu\lambda} A_{\mu} \p_\nu a_\lambda
\eeq
to the Lagrangian~\eqref{sEFT}, it leads to a description of the phase as not simply topologically ordered, but rather as a symmetry enriched topological (SET) phase, which is outside the scope of this review.

A clear manifestation of topological order in the Chern-Simons description of topological phases is that the degeneracy of ground states depends on the topological properties of the manifold the system lives on. Specifically, it is possible to show that while there are no topologically degenerate ground states if the system~\eqref{sEFT} is defined on a sphere, on a torus the GSD is $m$ (see Appendix~\ref{sGSD} for details). More generally, it can be shown that the ground state degeneracy of the $K$-matrix theory~\eqref{EFT} on an arbitrary Riemann surface of genus $g$ is $|$det$(K)|^g$~\cite{Wen95}.

Another characteristic feature of FQH states is the existence of chiral gapless edge excitations, which cannot be gapped by any local perturbations. Placing the system described by~\eqref{sEFT} on a semi-infinite plane, it is possible to derive the effective action governing these boundary modes (see Appendix~\ref{sEdge} for details). The resultant theory,
\beq
\label{edgeaction}
S_{\text{edge}} = \frac{m}{4\pi} \int d^2x\, \left(\p_t \varphi \p_x \varphi - v\p_x \varphi \p_x \varphi \right),
\eeq
is a conformal field theory (CFT), also known as a chiral Luttinger liquid or level $m$ Kac-Moody theory, which describes a chiral boson moving at a velocity $v$. Importantly, these one-dimensional boundary CFTs allow us to directly calculate the wave-function of the two-dimensional bulk, as was established in a landmark paper by Moore and Read~\cite{mooreread}. For instance, from the conformal blocks of the chiral CFT~\eqref{edgeaction} it is possible to derive the Laughlin wave-function for the $\nu = 1/m$ FQH state. 


\subsection{Generalized Chern-Simons Theories}
\label{highercs}

We now turn our attention towards two-dimensional higher rank generalizations of Chern-Simons theories, which host excitations with restricted mobility, including fractons. Here, we will focus on a specific chiral phase described by a rank 2 symmetric tensor gauge field---the ``scalar charge theory" in the taxonomy of Ref.~\cite{chiral}---to demonstrate the general phenomenology of such phases, with generalizations to different higher rank theories left for future work. 

We first consider a phase described by a rank 2 spatial symmetric tensor $A_{ij}$, with its canonically conjugate variable $E_{ij}$ playing the role of an electric field tensor. This theory, similar to the 3D rank 2 theory reviewed in Section~\ref{review}, is uniquely specified by a generalized Gauss' law which takes the form
\beq
\p_i \p_j E^{ij} = \rho,
\eeq 
for a scalar charge density $\rho$. The excitations carrying this charge obey two constraints, 
\beq
\int \rho = \text{const.},\quad \int \vec{x} \rho = \text{const.},
\eeq
corresponding to the conservation of charge and of dipole moment respectively. The fundamental charges of this theory are hence fractons, unable to move in any direction due to the dipole moment conservation law. Importantly, however, the dipolar bound states of this theory are completely mobile, possessing both longitudinal and transverse motion~\cite{genem}. The constraint in the low energy sector, $\p_i \p_j E^{ij} = 0$, leads to invariance under the gauge transformation 
\beq
A_{ij} \to A_{ij} + \p_i \p_j \alpha,
\eeq
for gauge parameter $\alpha(\vec{x},t)$ with arbitrary space-time dependence. The long-distance Hamiltonian consistent with this gauge structure is 
\beq
\label{ham}
\mathcal{H} = \frac{1}{2}E^{ij}E_{ij} + \frac{1}{2}B^i B_i,
\eeq
where the magnetic field is a vector quantity,
\beq
B^j = \epsilon^{ib} \p_i A_b^{\,\,\,j},
\eeq
and the constraint $\p_i \p_j E^{ij} = 0$ is implicitly assumed. The Hamiltonian~\eqref{ham} leads to a linearly dispersing gapless gauge mode with two polarizations. We also note that the magnetic flux vector $B^i$ in this case satisfies the constraint
\beq
\int x_i B^i = \text{const.},
\eeq
which implies that the magnetic fluxes are one-dimensional, with only transverse mobility. 

Following the discussion presented in Ref.~\cite{chiral}, we can also formulate this theory in terms of a Lagrangian by introducing a Lagrange multiplier field $\phi$ which imposes the Gauss' law constraint. The Lagrangian of this two-dimensional theory is
\beq
\label{bareaction}
\mathcal{L}_0(A_{ij},\phi) = \frac{1}{2} \left(\dot{A}_{ij} - \p_i \p_j \phi\right)^2 - \frac{1}{2}B^i B_i,
\eeq
which is gauge-invariant under the transformations
\beq
\begin{split}
A_{ij} &\to A_{ij} + \p_i \p_j \alpha, \\
\phi &\to \phi + \dot{\alpha},
\end{split}
\eeq
where $\alpha(\vec{x},t)$ has arbitrary space-time dependence. 

\subsubsection{Feeding the Gauge Field}
\label{gapped}

Having established the properties of the tensor gauge field with a Maxwell action, we now consider the effects of a Chern-Simons term, which we expect will gap the theory and perform some type of flux attachment.  As we discuss in the next section, a tensor Chern-Simons term attaches magnetic flux to dipoles.  First however, we verify that the Chern-Simons term fully gaps the gauge field, giving a gapped chiral phase of matter with fracton excitations.

We introduce a Chern-Simons term for the tensor gauge field $A_{ij}$ as:
\beq
\label{fullaction}
\mathcal{S}[A_{ij},\phi] = \mathcal{S}_0[A_{ij},\phi] + \mathcal{S}_{gCS}[A_{ij},\phi],
\eeq  
where we have added to the Lagrangian~\eqref{bareaction} a generalized Chern-Simons action
\beq
\begin{split}
\label{gencs}
\mathcal{S}_{gCS}[A_{ij},\phi] =& -\frac{\theta}{4\pi^2} \int d^3x\, \phi \epsilon^{bi} \p_i \p_j A_b^{\,\,\,j} \\
 &+ \frac{\theta}{8\pi^2}\int d^3x\, \epsilon^{bi} \dot{A}_{ij} A_b^{\,\,\,j} ,
\end{split}
\eeq
parametrized by $\theta$. We note that this Chern-Simons action can be derived as the boundary theory of a 3D higher rank tensor gauge field with a generalized ``E$\cdot$B" term parametrized by a coefficient $\theta$~\cite{chiral}. Here, we study the action~\eqref{gencs} in strictly two-spatial dimensions, observing that, unlike the vector Chern-Simons theory~\eqref{CS}, the generalized action~\eqref{gencs} does not describe a topological quantum field theory, as there does not appear to be any metric-independent formulation of this theory~\cite{chiral}.  

In order to understand the consequences of the Chern-Simons action for the gapless gauge modes, we decouple the symmetric gauge field $A_{ij}$ into its trace $\gamma$ and a symmetric traceless tensor $\tilde{A}_{ij}$
\beq
A_{ij} = \tilde{A}_{ij} + \gamma \delta_{ij}. 
\eeq
Substituting this into the action~\eqref{fullaction}, we obtain
\beq
\label{dcaction}
\begin{split}
\mathcal{S}[\tilde{A}_{ij},\gamma,\phi] &= \frac{1}{2}\int d^3x\, \left(\dot{\tilde{A}}_{ij} - \left(\p_i \p_j - \frac{1}{2}\delta_{ij} \p^2 \right)\phi \right)^2 \\
&- \frac{1}{2}\int d^3x\, \left(\p_i \gamma + \epsilon^{ij}\epsilon^{ab}\p_a \tilde{A}_{bj} \right)^2 \\
&+ \int d^3x\,\left(\dot{\gamma} - \frac{1}{2}\p^2 \phi \right)^2  + \mathcal{S}_{gCS}[\tilde{A}_{ij},\phi].
\end{split}
\eeq
Let us unpack this expression. First, we note that in terms of the separated fields $\tilde{A}_{ij}$ and $\gamma$, the action is invariant, up to a boundary term, under the gauge-transformation 
\beq
\begin{split}
\tilde{A}_{ij} &\to \tilde{A}_{ij} + \left(\p_i \p_j - \frac{1}{2} \delta_{ij} \p^2 \right)\alpha,  \\
\gamma &\to \gamma + \frac{1}{2}\p^2 \alpha,  \\
\phi &\to \phi + \dot{\alpha},
\end{split}
\eeq
where $\alpha(\vec{x},t)$ has arbitrary space-time dependence. Remarkably, we see that the trace component does not appear in the Chern-Simons action, which ostensibly gives a mass only to the traceless component of the gauge mode while leaving the trace component massless.

To decipher the fate of the traceless component, we focus on the terms in the action~\eqref{dcaction} containing $\gamma$,
\beq
\begin{split}
\mathcal{S}[\tilde{A}_{ij},\gamma,\phi] &=  \int d^3x\,\left(\dot{\gamma} - \frac{1}{2}\p^2 \phi \right)^2 \\
&- \frac{1}{2}\int d^3x\, \left(\p_i \gamma + \epsilon^{ij}\epsilon^{ab}\p_a \tilde{A}_{bj} \right)^2 + \mathcal{S}_2[\tilde{A}_{ij},\phi]
\end{split}
\eeq
where $\mathcal{S}_2$ contains the remaining terms in the action. Re-parametrizing $\tilde{A}_{ij}$ in terms of an effective gauge field
\beq
\Gamma^i = \epsilon^{ij} \epsilon^{ab} \p_a \tilde{A}_{bj}, 
\eeq 
we can re-write Eq.~\eqref{dcaction} as
\beq
\begin{split}
\mathcal{S}[\tilde{A}_{ij},\gamma,\phi] &=  \int d^3x\,\left[\left(\dot{\gamma} - \frac{1}{2}\p^2 \phi \right)^2 - \frac{1}{2}\left(\p_i \gamma + \Gamma_i \right)^2\right] \\
&+ \mathcal{S}_2[\tilde{A}_{ij},\phi].
\end{split}
\eeq
Once written in this form, it is clear that the trace mode $\gamma$ couples to the effective gauge field $\Gamma_i$ in a manner redolent of the coupling between a superfluid phase $\varphi$ and an ordinary vector potential $A_\mu$: $(\p_\mu \varphi - A_\mu)^2$. By analogy with the case of a superfluid, where the vector potential ``eats" the gapless Goldstone mode, we thus expect that the gapless trace mode will get eaten by the effective gauge field $\Gamma$. This can be seen explicitly by making a gauge transformation
\beq
\tilde{A}_{ij} \to \tilde{A}_{ij} + \left(\p_i \p_j - \frac{1}{2}\delta_{ij} \p^2 \right)\alpha, \quad \phi \to \phi + \dot{\alpha},
\eeq
with $\alpha(\vec{x},t)$ such that
\beq
\frac{1}{2}\p^2 \alpha = \gamma.
\eeq
While $\mathcal{S}_2[\tilde{A}_{ij},\phi]$ is invariant under such a transformation, the magnetic flux vector and effective gauge field transform as
\beq
\begin{split}
B^j &\to \tilde{B}^j = \epsilon^{ib} \p_i \tilde{A}_b^{\,\,\,j} ,\\
\Gamma_i &\to \Gamma_i - \frac{1}{2}\p_i \p^2 \alpha = \Gamma_i - \p_i \gamma,
\end{split} 
\eeq 
thereby completely eliminating the trace mode $\gamma$ from the theory. Surprisingly, we have found that the gauge field eats its own trace component, leading us to christen this higher rank tensor gauge field an ``ouroboros" gauge field. We note that similar behavior may be displayed by non-Abelian vector gauge theories in the context of SU(2) spin liquids~\cite{wenbook}. 

Since $\gamma$ is a compact field, we should, in principle, account for the presence of vortices as is often done when studying ordinary superfluids. Specifically, on splitting $\gamma$ into a regular part $\gamma_r$ and a singular part $\gamma_s$, only the regular part would get absorbed into the effective gauge field $\Gamma_i$, with $\gamma_s$ describing gapped vortices with short-range interactions. We can then also imagine integrating out the gapped gauge mode, leading to an effective action for these vortices, which would take the form of a conventional vector Chern-Simons theory. We leave a detailed discussion of these vortices for future work. 

The preceding discussion establishes that the higher rank Chern-Simons term leads to a completely gapped phase, with the low energy physics of this phase described by a traceless symmetric rank 2 field $\tilde{A}_{ij}$~\footnote{Since the trace mode disappears from the theory, it is natural to ask whether we could simply start from the theory of a traceless symmetric rank 2 tensor. However, the particle structure for such a theory would differ from the one studied here and must thus be considered separately.}. The effective action describing the long-distance physics of this gapped phase is hence the generalized Chern-Simons term
\beq
\begin{split}
\label{geneft}
\mathcal{S}_{gCS}[\tilde{A}_{ij},\phi] =& -\frac{\theta}{4\pi^2} \int d^3x\, \phi \epsilon^{bi} \p_i \p_j \tilde{A}_b^{\,\,\,j} \\
 &+ \frac{\theta}{8\pi^2}\int d^3x\, \epsilon^{bi} \dot{\tilde{A}}_{ij} \tilde{A}_b^{\,\,\,j} ,
\end{split}
\eeq
since it has one fewer derivative than the Maxwell-like term in the action. As mentioned earlier, in principle we should also include the gapped vortices of the trace mode $\gamma$, but we will focus here on the physics captured by the higher rank Chern-Simons term~\eqref{geneft}, which is gauge-invariant (on a closed manifold) under
\beq
\tilde{A}_{ij} \to \tilde{A}_{ij} + \left(\p_i \p_j - \frac{1}{2}\delta_{ij} \p^2 \right)\alpha, \quad \phi \to \phi + \dot{\alpha},
\eeq
for arbitrary $\alpha(\vec{x},t)$. 

Similarly to the quantization of the level $k$ of a vector Chern-Simons theory, the coefficient $\theta$ of the action~\eqref{geneft} is also quantized in units of $2\pi$ (see Appendix~\ref{quant} for a derivation),
\beq
\theta = 2 \pi k,\quad k\in\mathbb{Z}.
\eeq 
We will henceforth refer to $k$ as the level of the higher rank tensor Chern-Simons theory, which we will now demonstrate describes an emergent integer quantum Hall state of mobile dipoles at the filling fraction $\nu = k$. We note that, similarly to the CS theory describing the quantum Hall effect, we will see that when $k \in \mathbb{Z}$, the generalized CS term~\eqref{geneft} describes an IQH state of dipoles while for fractional values, it is describing an FQH state.

\subsubsection{Integer quantum Hall state of Dipoles}
\label{dipoleqh}

So far, we have established that the addition of the higher rank Chern-Simons term leads to a fully gapped phase, where the trace mode of the symmetric tensor $A_{ij}$ has been eaten by the traceless mode $\tilde{A}_{ij}$. With the gauge sector of the theory fixed, we now examine the particle content of the low-energy theory~\eqref{geneft}. 

The Lagrange multiplier field $\phi$ constrains the low-energy sector of this theory,
\beq
-\frac{k}{2\pi} \epsilon^{bi}\p_i\p_j \tilde{A}_b^{\,\,\,j} = \frac{k}{2\pi} \p_j \tilde{B}^j = 0,
\eeq
where $\tilde{B}^j$ is the magnetic flux vector. More generally, allowing for the appropriate gauge charges coupled to the Chern-Simons field, 
\beq
\label{fluxattach}
\rho = -\frac{k}{2\pi} \epsilon^{bi}\p_i\p_j \tilde{A}_b^{\,\,\,j}.
\eeq
It can readily be checked that the excitations carrying this charge obey the constraints
\beq
\int \rho = \text{const.},\quad \int \vec{x} \rho = \text{const.}, \quad \int x^2 \rho = \text{const.,}
\eeq
corresponding to the conservation of charge, dipole moment, and a specific component of the quadrupole moment. The fundamental charges are thus fractons, while the dipolar bound states are only mobile in the direction transverse to their dipole moment. 

While it seems surprising that the dipoles, which were fully mobile in the absence of the Chern-Simons term, now have restricted mobility, this restriction arises naturally as a consequence of flux-attachment~\eqref{fluxattach},
\beq
\rho = \frac{k}{2\pi} \p_j \tilde{B}^j,
\eeq
which indicates that the Chern-Simons term binds a flux $2\pi/k$ to each dipole. Since the magnetic flux vector in this theory is one-dimensional, it follows that dipole excitations in the gapped deconfined phase inherit the restricted mobility of the fluxes. 

\begin{figure}[b!]
 \centering
 \includegraphics[width=0.5\textwidth]{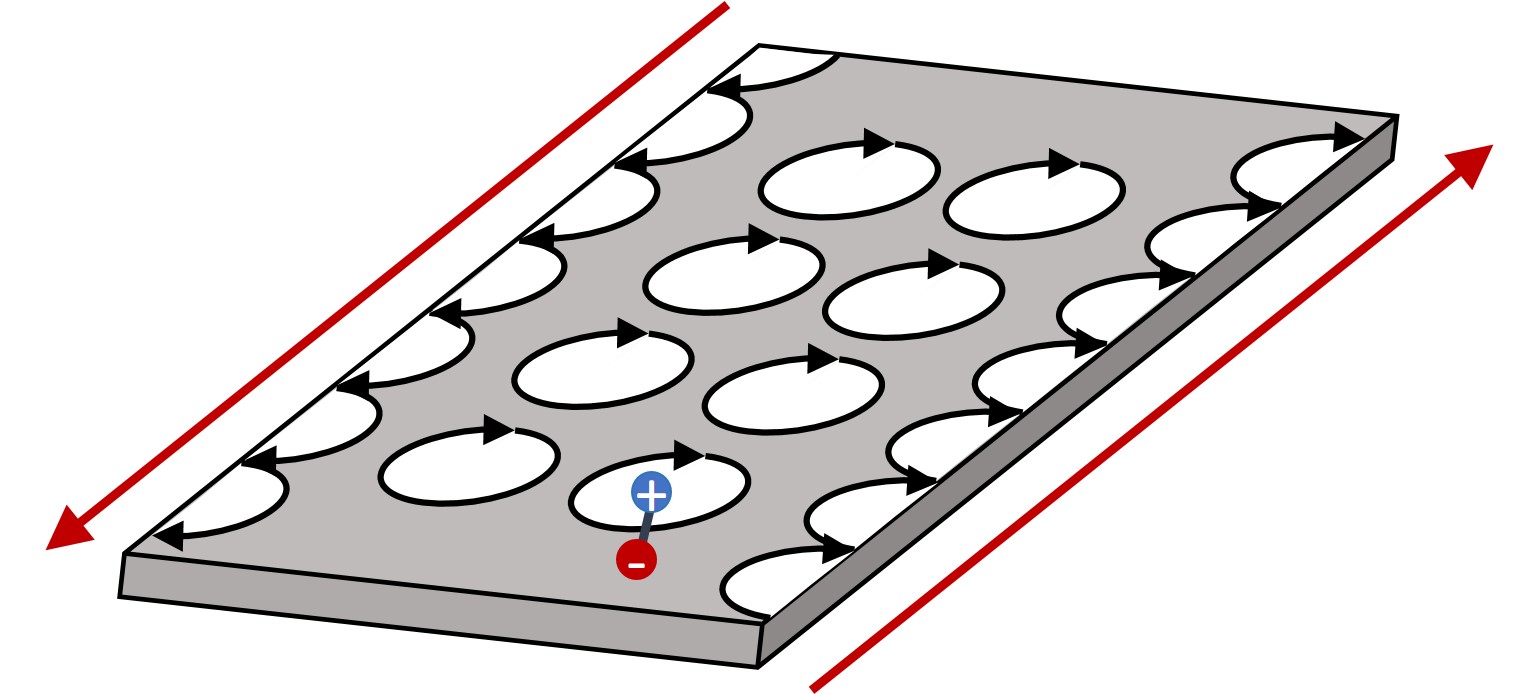}
 \caption{Fully mobile dipoles, indicated by the pair of connected spheres, in the presence of an emergent finite background magnetic field perform cyclotron motion. In the presence of a boundary, these skipping orbits result in chiral dipolar edge currents.}
 \label{fig:orbits}
 \end{figure}

The physics of this gapped phase can be understood through a simple semi-classical picture, where fully mobile dipoles at some finite density move in the presence of an emergent finite background magnetic field. In analogy with electrons in a perpendicular magnetic field, these dipoles move in quantized circular orbits and perform cyclotron motion, as depicted in Figure~\ref{fig:orbits}. In this context, the integer $k$ has a natural interpretation as the number of filled Landau levels $\nu$ occupied by the mobile dipoles, and this phase hence corresponds to an emergent integer quantum Hall phase of dipoles. Since we are attaching an integer amount of flux to the dipoles, which are originally mobile, there is no fractionalization of charge or statistics of the deconfined quasi-particles. Instead, a striking new feature of the generalized Chern-Simons theory considered here is the fractionalization of mobility of the underlying dipoles forming the state, resulting in deconfined dipolar excitations with only one-dimensional motion allowed.

Pushing on the semi-classical picture further, we expect that the circular orbits will reduce to skipping orbits in the presence of a boundary, leading to a chiral dipolar current propagating along the edge. Indeed, as we will show later (see Section~\ref{edge}), the presence of these boundary modes can be established directly by placing the tensor Chern-Simons theory~\eqref{geneft} on an open manifold.

Besides these boundary modes, we can also characterize this phase through its generalized ``Hall" response, which follows from varying the action~\eqref{geneft} with respect to $\tilde{A}_{ij}$,
\beq
\langle J^{ij} \rangle = \frac{k}{4\pi} \left(\epsilon^{ib} \dot{\tilde{A}}_b^{\,\,\,j} + \epsilon^{jb}\dot{\tilde{A}}_b^{\,\,\,i} \right) = \frac{k}{4\pi}\left(\epsilon^{ib} E_b^{\,\,\,j} + \epsilon^{jb} E_b^{\,\,\,i} \right).
\eeq
We can thus define a ``Hall" conductivity which is given by 
\beq
\sigma^{ijkl} = \frac{k}{4\pi}\left(\epsilon^{ik} \delta^{jl} + \epsilon^{jk}\delta^{il} \right).
\eeq
The only non-trivial components of this tensor are
\beq
\sigma^{xyyy} = \sigma^{yxyy} = -\sigma^{yxxx} = -\sigma^{xyxx} = \frac{k}{4\pi},
\eeq
and 
\beq
\sigma^{xxyx} = -\sigma^{yyxy} = \frac{k}{2\pi}.
\eeq
Unlike the Hall response in an integer quantum Hall state of electrons, which represents the response of the system to an externally varying electric field, here the conductivity tensor $\sigma^{ijkl}$ encodes the response to the internal emergent tensor field $E_{ij}$.

\subsubsection{Fractional quantum Hall analogues}
\label{dipolefqh}

Thus far, we have focused on the case where $k$ is restricted to take integer values, which is reflected in the lack of fractional charges in the system and corresponds to a generalized Hall response which is an integer multiple of a fundamental constant. In analogy with the fractional quantum Hall effect, we can also consider fractional values of the level $k$, which will lead to deconfined fractons with fractional charges and a fractional generalized Hall response. 

In order to study the fractional generalizations of the tensor Chern-Simons theory, we introduce a rank 2 symmetric traceless gauge field $a_{ij}$. In analogy with the Chern-Simons description of FQH states (see Section~\ref{csreview}), the higher rank Chern-Simons theory of this compact U(1) statistical gauge field will capture the long-distance physics of a fracton phase with fractionalized charges. The effective low-energy theory for an emergent fractional quantum Hall state of dipoles at filling fraction $\nu = k = 1/m$ ($m\in\mathbb{Z}$) is described by
\beq
\label{fqhaction}
\begin{split}
\mathcal{S}[\tilde{A}_{ij},\phi ; a_{ij},\chi] = \mathcal{S}_{gCS}[a_{ij},\chi] + \mathcal{S}_{c}[\tilde{A}_{ij},\phi ; a_{ij},\chi] \\
+ \mathcal{S}_{Max}[\tilde{A}_{ij},\phi].
\end{split}
\eeq
The first term in the action is the generalized Chern-Simons term at level $m$ for the statistical gauge field $a_{ij}$ and a Lagrange multiplier field $\chi$, which enforces a low-energy constraint on the theory, leading to fractionalized excitations with restricted mobility.  The last term is the bare Maxwell action of the original tensor gauge field of the emergent fractons.

The middle term
\beq
\begin{split}
\mathcal{S}_c &= \frac{1}{2\pi} \int d^3x\, \phi\,\epsilon^{bi} \p_i \p_j a_b^{\,\,\,j} + \frac{1}{2\pi}\int d^3x\,\chi\,\epsilon^{bi}\p_i \p_j \tilde{A}_b^{\,\,\,j} \\
& - \frac{1}{2\pi} \int d^3x\,\epsilon^{bi} \dot{\tilde{A}}_{ij} a_b^{\,\,\,j},
\end{split}
\eeq
describes the coupling between the emergent electro-magnetic field $\tilde{A}_{ij}$ and $a_{ij}$; this is the analogue of the mixed Chern-Simons term~\eqref{regcoup} in the vector Chern-Simons theory, which couples the statistical gauge field to the physical gauge potential.  This coupling term will be our primary focus from hereon.  Though the field $\tilde{A}_{ij}$ is technically dynamical, we will now treat it as an effective static background source.  This is completely analogous to the normal treatment of quantum Hall states, where the dynamical electromagnetic field is treated as an effective background.

That the action~\eqref{fqhaction} is indeed the correct low energy description of an emergent FQH state can be verified directly by integrating the massive gauge field $a_{ij}$ out of the theory---a few lines of algebra will lead to the action~\eqref{geneft} with coefficient $\theta = -2\pi/m$, where $2\pi m$ flux is attached to each dipole. Note that while we could have added a generalized Chern-Simons term for $\tilde{A}_{ij}$ to~\eqref{fqhaction}, this merely shifts the generalized Hall response by an integer. Thus, setting the coefficient of this term to zero corresponds to working in the lowest Landau level. 

Alternatively, we could have started from the effective theory for $\tilde{A}_{ij}$ and introduced a finite density of dipoles $\rho$ in addition to a current $J_{ij}$, by including source terms
\beq
\mathcal{L}_{\text{source}} = -\phi \rho - \tilde{A}_{ij}J^{ij},
\eeq
to the action~\eqref{geneft}. As discussed in Section~\ref{review}, these satisfy the generalized continuity equation
\beq
\dot{\rho} + \p_i \p_j J^{ij} = 0,
\eeq
which is solved by taking
\beq
\rho = \frac{1}{2\pi}\epsilon^{ib}\p_i\p_j a_b^{\,\,\,j},\quad J^{ij} = \frac{\epsilon^{bi}}{2\pi}\left(\dot{a}_b^{\,\,\,j} + \p_b \p^j \chi\right), 
\eeq
for arbitrary $\chi$. Inserting these into $\mathcal{L}_{\text{source}}$ and adding a generalized Chern-Simons term for the $a_{ij},\chi$ fields leads precisely to the action~\eqref{fqhaction} for an emergent FQH fracton phase. Note that similar arguments are used to arrive at a Chern-Simons effective description for conventional FQH states of electrons~\cite{wenbook}.

The field $\chi$ plays the role of a Lagrange multiplier and places a low-energy constraint on the gauge charges of the statistical gauge field $a_{ij}$. Specifically, these charges are fractons, while their dipolar bound states are one-dimensional, with mobility only in the direction transverse to their dipole moment. We can also explicitly see that the quasi-particle excitations carry fractionalized charge by introducing an excitation that carries gauge charge $q$ under $a_{ij}$. This is achieved by adding a term
\beq
\delta \mathcal{L} = q \chi\,\delta(\vec{x} - \vec{x}_0)
\eeq
to the action~\eqref{fqhaction}. Varying the action with respect to $\chi$ leads to the equation of motion
\beq
\label{eom}
\rho = \frac{1}{2\pi m}\p_j B^j + \frac{q}{m} \delta(\vec{x} - \vec{x}_0),
\eeq  
which explicitly demonstrates that the filling fraction of the underlying dipoles forming the state is $\nu = 1/m$ and that the one-dimensional dipolar excitations carry fractional charge $q/m$. Following the arguments for electronic FQH states~\cite{wenbook}, it is now easy to show that the excitations in this theory have fractional statistics~\footnote{While the dipolar excitations are one-dimensional, there still exists a well-defined notion of statistics for sub-dimensional particles. This is discussed in detail in the context of discrete fracton models~\cite{han}.}. In particular, we can define a ``braiding" process along intersecting lines for a $q_1$ dipole with a $q_2$ dipole which has dipole moment orthogonal to that of the $q_1$ dipolar quasi-particle.  Such a process will induce a phase
\beq
\theta_{12} = \frac{2\pi}{m}q_1 q_2.
\eeq
Additionally, we can infer from the equation of motion~\eqref{eom} that a quasi-particle carrying $m$ units of $a_{ij}$ charge corresponds to a dipole excitation forming the FQH liquid. We can identify the excitations carrying $m$ units of the $a$ charge with the bosonic (fermionic) dipoles forming the underlying FQH state when $m$ is even (odd). 

As discussed above, for cases where $k=1/m$, the excitations exhibit fractionalized charge and statistics, and our experience with FQH states of electrons would lead us expect a constant ground state degeneracy on a torus. There is, however, a crucial difference between electrons and the dipolar bound states---the dipoles have mobility only transverse to their dipole moment. Imagine placing the system on an $R_x \times R_y$ periodic lattice with lattice spacing $a=1$. Now consider the Wilson-line operator that creates a pair of dipoles at spatial position $(x,y)=(0,0)$, wraps one of the dipoles along a non-contractible cycle in the $y$-direction, and annihilates it with its partner to return the system to the vacuum state. If the dipoles were fully mobile, this operator could be locally deformed into one which initially creates the pair at $(x,y)=(1,0)$. Due to the one-dimensional nature of the dipoles, it appears that these operators may no longer be continuously deformed into one another. While this may lead one to expect a ground state degeneracy which grows exponentially with system size, there in fact exist a sub-extensive number of relations between certain products of the Wilson string operators which reduce the degeneracy to a constant\footnote{A similar situation arises in discrete fracton models such as the X-Cube model, where certain relations between products of Wilson string operators reduce the ground state degeneracy from $\sim 2^{L^2}$ down to $\sim2^L$~\cite{slagle2}}. Indeed, a detailed derivation (see Appendix~\ref{GSD}) shows that the ground state degeneracy for the generalized Chern-Simons theory $\mathcal{S}_{gCS}[a_{ij},\chi]$ for the statistical field $a_{ij}$ at level $m$ (where $k = 1/m$ is the filling fraction) is a constant,
\beq
\text{GSD} = 2m,
\eeq
where the factor of $m$ arises as a consequence of the fractional statistics (the factor of two stems from having two species of dipoles).

Having established this formalism, it is now tempting to generalize this construction to dipolar analogues of hierarchical quantum Hall states, described by a $K$-matrix and a multiplet of tensor gauge fields $a_{ij}^{\,I}$, or of non-Abelian quantum Hall states, such as the Moore-Read Pfaffian state~\cite{mooreread,fradkin}. While such generalizations appear fairly straightforward to construct, we will leave this for future work and focus instead on the novel boundary theories of these generalized Chern-Simons theories. 

\subsubsection{Chiral Edge Modes}
\label{edge}

From the semi-classical picture for the emergent quantum Hall state of dipoles, where a finite density of dipoles responds to a finite background magnetic field, we expect that the system will host chiral modes localized at spatial boundaries. This is illustrated schematically in Figure~\ref{fig:orbits} for the quantum Hall state of dipoles, described by the higher rank Chern-Simons theory with level $m \in \mathbb{Z}$.

Here, we will show the existence of chiral edge modes explicitly, working with the generalized Chern-Simons theories for dipolar FQH states at filling fraction $1/m$. Indeed, in analogy with vector Chern-Simons theories describing FQH states of electrons, we expect that the chiral higher rank Chern-Simons theories also exhibit a chiral anomaly, a direct manifestation of which are gapless chiral edge modes. Note that the case $m=1$ describes dipoles in a completely filled lowest Landau level, i.e., an IQH state of dipoles. 

Consider the higher rank Chern Simons term describing dipoles at filling fraction $\nu = k = 1/m$,  
\beq
\label{genact}
\mathcal{S}[a_{ij},\chi] = -\frac{m}{2\pi} \int d^3x\,\chi\epsilon^{bi}\p_i \p_j a_b^{\,\,\,j} + \frac{m}{4\pi} \int d^3x\,\epsilon^{bi} \dot{a}_{ij}a_b^{\,\,\,j},
\eeq
where $a_{ij}$ is a compact U(1) symmetric traceless tensor of rank 2. Under a gauge transformation 
\beq
\begin{split}
a_{ij} &\to a_{ij} + \left(\p_i \p_j - \frac{1}{2}\delta_{ij} \delta^2 \right)\alpha, \\
\chi & \to \chi + \dot{\alpha},
\end{split}
\eeq
this action is invariant only up to a boundary term,
\beq
\mathcal{S}[a_{ij},\chi] \to \mathcal{S}[a_{ij},\chi] + \frac{m}{4\pi} \int d^3x\, \p_i \left(\epsilon^{bi} \p_j \dot{\alpha}\,\p_b \p^j \alpha \right).
\eeq
To derive the action for the boundary degrees of freedom, we fix the gauge $\chi = 0$ in the bulk such that the constraint imposed by the gauge-fixing condition remains
\beq
\epsilon^{bi} \p_i\p_j a_b^{\,\,\,j} = 0.
\eeq
This constraint can be solved in terms of the field $\varphi$
\beq
\label{edgesol}
a_b^{\,\,\,j} = \epsilon^{ij}\p_i \p_b \varphi.
\eeq
Note that unlike the edge theory of a vector CS theory (see Appendix~\ref{sEdge}), which is described in terms of compact scalar field, here $\varphi$ has dimensions of length ($[\varphi] = L^1$) and it is instead $\p \varphi$ which is a compact dimensionless field. In addition, since $a_{ij}$ is a symmetric tensor, $\varphi$ must be a solution of the two-dimensional Laplace equation
\beq
\p^2 \varphi = 0.
\eeq
Tracelessness of $a_{ij}$ is automatically satisfied by \eqref{edgesol}.
\begin{figure}[b!]
 \centering
 \includegraphics[width=0.45\textwidth]{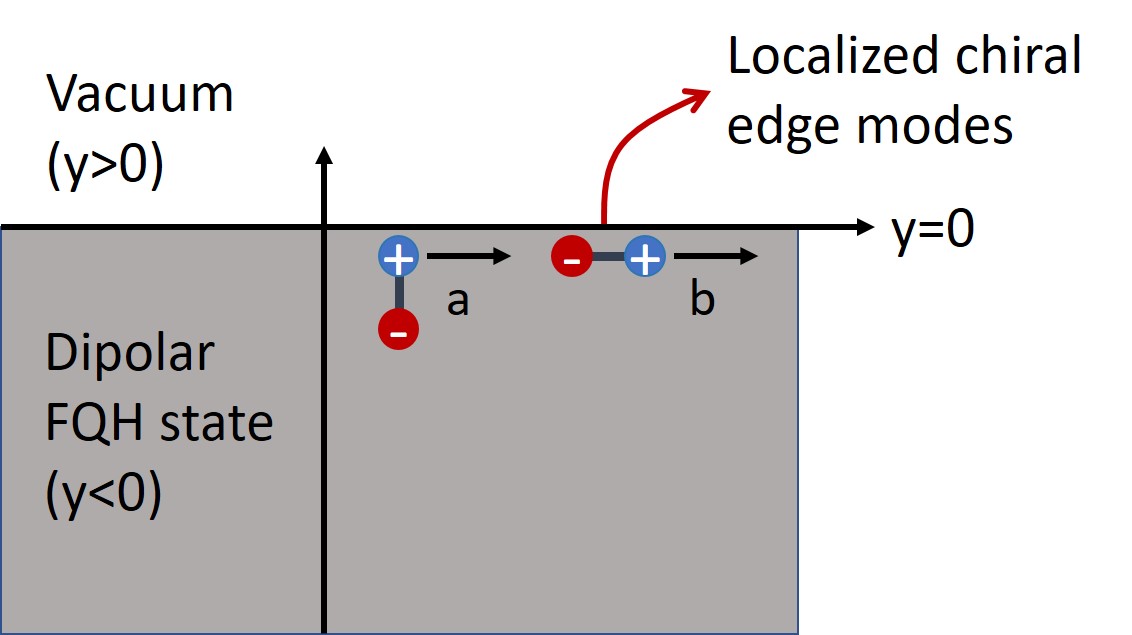}
 \caption{Semi-infinite geometry for studying the edge physics of the higher rank Chern-Simons theory. The boundary at $y=0$ separates the dipolar FQH state from the vacuum. The boundary hosts two independent co-propagating gapless chiral modes, which correspond to the motion of dipolar bound states in two distinct orientations, $a$ and $b$, along the edge.}
 \label{fig:edge}
 \end{figure}

For concreteness, we consider the semi-infinite geometry depicted in Figure~\ref{fig:edge}, which has a spatial boundary at $y=0$ between the dipolar FQH state and vacuum. Inserting the solution~\eqref{edgesol} into the action~\eqref{genact}, we obtain the edge action
\beq
\begin{split}
\mathcal{S}_{gCS} &= \frac{m}{4\pi}\int_{y=0} dxdt\, \left(\p_t \p_x \varphi\,\p_x\p_x \varphi \right) \\
&+ \frac{m}{4\pi}\int_{y=0} dxdt\, \left(\p_t \p_y \varphi\,\p_x\p_y \varphi \right).
\end{split}
\eeq
Since $\varphi$ and $\p_y \varphi$ can be varied independently on the boundary, we have thus found two modes propagating along the boundary.

Physically, it is natural to expect two distinct boundary modes corresponding to the two linearly independent dipole orientations. This is depicted schematically in Figure~\ref{fig:edge}. In particular, this figure illustrates the distinct nature of these two boundary modes. Consider first dipoles with their dipole moment oriented perpendicular to the boundary and propagating transversely to their dipole moment. Labelled by $a$ in Figure~\ref{fig:edge}, the motion of these dipoles along the boundary looks identical to that of a usual charged particle. The boundary action for this mode should hence be identical to that of a chiral Luttinger liquid of electrons. Indeed, when re-phrased in terms of a new compact scalar field $\xi \equiv \p_y \varphi$, the boundary action for $\xi$ becomes
\beq
\mathcal{S}_{\text{edge}}[\xi] = \frac{m}{4\pi}\int_{y=0} dx dt\, \left(\p_t \xi \,\p_x \xi \right),
\eeq
which is precisely the chiral CFT describing the boundary of a conventional FQH state (see Section~\ref{csreview}). We can thus introduce a velocity for this field by adding a term $-v \p_x \xi\,\p_x \xi$ to the action. Given the extensive literature on edge theories of quantum Hall states (see e.g.,~\cite{wenbook}), we will not investigate this mode further. 

In addition to the conventional mode described by $\xi$, the boundary of a rank 2 Chern-Simons theory hosts an additional chiral edge mode, with the low-energy behavior governed by
\beq
\label{gET}
\mathcal{S}_{gET}[\varphi] = \frac{m}{4\pi}\int dxdt\, \left(\p_t \p_x \varphi\,\p_x\p_x \varphi \right),
\eeq
where ``gET" denotes a ``generalized edge theory," and it is implicit that this theory lives on the boundary $y=0$. As depicted in Figure~\ref{fig:edge}, this mode ($b$) corresponds to the motion of dipoles with their dipole moment oriented parallel to the boundary, with longitudinal motion along the boundary. Since dipole moment is conserved, these excitations are constrained to move in dipolar bound pairs even along the boundary. As a consequence of this, we expect the low energy theory describing these edge modes to obey an additional constraint besides charge conservation. We will now explicitly demonstrate this by working directly with the action~\eqref{gET}. 

As noted earlier, the field $\varphi$ has dimensions of length and so under scale transformations of the coordinates
\beq
\sigma^a \to \lambda \sigma^a
\eeq
transforms as
\beq
\varphi(\sigma) \to \varphi(\lambda^{-1} \sigma),
\eeq
leaving the action~\eqref{gET} invariant under scale transformations. Indeed, due to the unconventional scaling of the field $\varphi$, the action~\eqref{gET} is a conformally invariant field theory despite initial appearances. 

In addition to the term~\eqref{gET}, we can also add non-universal energetic terms to the action. Following standard procedure, we identify $\p_x \p_x \p_x \varphi$ as the field  canonically conjugate to $\varphi$, from which we can derive the commutation relations
\beq
[\p_x \varphi(x),\p_x \varphi(x')] = -\frac{i \pi}{m} \text{sgn}(x - x'),
\eeq
establishing both $\varphi$ and $\p_x \varphi$ as non-local fields. Thus, adding the lowest order spatial derivative term, consistent with locality, we find the low-energy theory for the field $\varphi$
\beq
\mathcal{S}_{gET}[\varphi] = \frac{m}{4\pi}\int dxdt\, \left(\p_t \p_x \varphi\,\p_x\p_x \varphi - v \p_x \p_x \varphi\,\p_x\p_x \varphi\right),
\eeq
which describes a chiral gapless mode with dispersion $\omega = v k$, where $v$ is some non-universal velocity determined by the microscopic details of the edge. Note that the velocity of this mode will generically be distinct from that of the $\xi$ edge mode, although the two modes will co-propagate, resulting in a thermal Hall coefficient which is twice that of conventional FQH states.

Given the novelty of the generalized edge theory, we leave a thorough discussion of its properties to future work. Instead, we focus on a remarkable new property of this theory, absent in boundary theories of vector Chern-Simons theories. Specifically, we observe that the density of dipoles at the boundary is defined by 
\beq
\rho = \frac{1}{2\pi}\p_x \p_x \varphi,
\eeq
as evinced by the commutation relations
\beq
\begin{split}
[\rho(x),\p_x \varphi(x')] &= -\frac{i}{m}\delta(x - x'),\\
[\rho(x),\rho(x')] &= \frac{i}{2\pi m} \p_x \delta(x-x').
\end{split}
\eeq
Due to the additional spatial derivative present in $\rho$, the dipolar density at the boundary satisfies the constraints
\beq
\int \rho = \text{const.},\quad \int x \rho = \text{const.},
\eeq
corresponding to conservation of charge and of centre of mass, the latter of which we anticipated in our earlier discussion. We have thus found that the boundary theory of the generalized Chern-Simons theory hosts gapless excitations with centre of mass conservation i.e., there exist gapless fracton excitations at the edge of a higher rank Chern Simons theory in addition to fully mobile one-dimensional dipolar bound states. Since chirality guarantees the absence of back-scattering terms, regardless of the interactions present at the boundary, we expect that both of these edge modes will be robust against arbitrary weak perturbations. However, given the unusual nature of the generalized edge theory, we leave a careful analysis of stability to future work.

Although we have focused on a semi-infinite planar geometry here, in principle we could consider other finite samples as well. Interestingly, we expect that on a square geometry (or any sample with sharp corners) there will exist finite-energy localized modes at the corners, owing to the one-dimensional nature of the dipolar excitations. While this behavior is distinct from the zero-energy protected corner modes recently discovered in higher-order topological insulators~\cite{teohughes,song,schindler,langbehn,hughes2,hughes3,fangfu}, the existence of such corner-localized modes opens up the possibility of finding protected zero-energy corner modes or hinge states in fracton models. We leave a detailed study of the generalized edge theory and its localized modes on general finite samples for the future.

As a final remark, we note that the generalized edge theory~\eqref{gET} provides us with another window into the emergent FQH state present in the bulk. Specifically, since the edge theory is Gaussian, the correlation function for the $\p_x \varphi$ fields can be calculated exactly and takes the suggestive form
\beq
\langle \p_x\varphi(x) \p_x \varphi(0)\rangle = - \frac{1}{m} \log(x - vt),
\eeq
where $m$ is the inverse filling fraction $\nu$ of the dipolar bound states. Since the boundary theory is conformal, we conjecture that the bulk-boundary correspondence for vector Chern-Simons theories holds also for the higher rank chiral theory studied here. Hence, from the correlator of the vertex operators $e^{i m \p_x \varphi}$ (which correspond to creation operators of elementary dipoles at the boundary),
\beq
\langle :e^{i m \p_x \varphi(x)}:\, :e^{i m \p_x \varphi(0)}: \rangle \sim \frac{1}{(x - vt)^m},
\eeq
we should be able to construct the bulk wave-function for the emergent dipolar degrees of freedom, which in this case will take the form of a Laughlin state at filling fraction $\nu = 1/m$. We stress that while this is consistent with the picture developed here, the real object of interest would be the wave-function of the underlying degrees freedom from which the higher rank tensor gauge field emerges.

Nonetheless, we have discovered an intriguing connection between ``fracton" phases and quantum Hall physics, in that chiral higher rank tensor gauge fields with a finite density of mobile dipoles form an emergent fractional quantum Hall phase, albeit with novel ground state degeneracies and edge physics.  We have identified chiral gapless edge theories which transport dipoles along the edge of the system.  As emphasized throughout, these dipoles do not carry physical electric charge, so there is no Hall response of the electric current.  Nevertheless, the dipoles carry energy, and in the spin liquid context, will carry spin.  A spin liquid supporting a dipolar quantum Hall phase will therefore be characterized by robust thermal and spin Hall responses, which will be useful for identifying these phases experimentally.

While we have focused only on a specific higher rank tensor gauge theory (the ``scalar charge" theory) in this work, a larger class of chiral two-dimensional phases hosting subdimensional fractons has been recently uncovered~\cite{chiral}. It would certainly be of interest to generalize our analysis here to these other cases, with the aim of developing a more general framework for understanding these higher rank chiral phases. For instances, in the ``traceless scalar charge" theory~\cite{genem}, flux is attached directly to the fractons and not their dipolar bound states, which are mobile only in the direction transverse to their dipole moment even prior to flux attachment. The physics of a finite density of fractons in a background emergent magnetic field remains to be understood, with simple analogies to quantum Hall states ruled out due to immobility of fractons. We leave such questions to future work.


\section{Conclusion}

In this paper, we have studied systems with a finite density of fractons or their dipolar bound states, mapping out some of the interesting phases in which this emergent fractonic matter can exist.  In so doing, we have initiated the study of ``condensed matter" of fractons.  We have uncovered a cornucopia of new phases including `fractonic' microemulsions, Fermi liquids, and quantum Hall states, as well as new finite temperature phase transitions. Of course, there remain numerous open directions.  In principle, the mobile dipoles can be placed into any phase of matter accessible to conventional bosons or fermions, and we have only studied a small sampling of possible phases.  As an example, it would be interesting to investigate topologically ordered states of dipoles.  There are also still many generalizations to explore within the dipolar Fermi liquid and quantum Hall frameworks, as we have discussed in the main text.  Furthermore, it is far from obvious that such analogues of conventional phases provide an exhaustive account of dipolar phases of matter.  There may be intrinsically new fractonic phases of matter with no natural analogue in conventional condensed matter, which is an intriguing possibility. We leave further investigation of the condensed matter physics of fractonic matter to future work.

\section*{Acknowledgments}

We thank S.A. Parameswaran for key conversations that inspired this work. We also acknowledge useful conversations with Kevin Slagle, Yang-Zhi Chou, Victor Gurarie, Mike Hermele, Sheng-Jie Huang, William Jay, Sergej Moroz, Olexei Motrunich, Albert Schmitz, Han Ma, and T. Senthil. We also thank our anonymous PRB referee who pointed out an error in an earlier version of the draft. M. P. is supported partially by NSF Grant 1734006 and partially by a Simons Investigator Award to Leo Radzihovsky.  A. P. and R. N. are supported by the Sloan Foundation through a Sloan Reseach Fellowship and by the U.S. Air Force Office of Sponsored Research (AFOSR) under the Young Investigator Program, award number FA9550-17-1-0183.

\appendix


\section{Screening in the Dipolar Fermi Gas}

We will here treat the different types of screening that occur in the single-species dipolar Fermi gas, adapting the techniques of Ref.~\cite{screening}.

\subsection{Dipole-Dipole Screening}

We first investigate the ability of the dipoles to screen their interactions with each other.   We emphasize that the following is an electrostatic calculation which will not capture e.g. dynamical screening properties of the phase, but an electrostatic calculation of this form is sufficient for our present purposes. 

Let us consider a system with an equilibrium density of fermionic dipoles, all with the same orientation $p^i$, forming a Fermi surface.  We then consider adding a static ``test dipole" in the $p^i$ direction and seeing how the system responds.  In the presence of a perturbing scalar potential $\phi$, the potential energy felt by a dipole is $p^i\partial_i\phi$~\cite{genem}.  Correspondingly, the induced dipole moment density is $-p^j(gp^i\partial_i\phi)$, where $g$ is the density of states of the dipoles at the Fermi surface.  Using the form of the potential created by a dipole, $\phi_p (r) = (p\cdot r)/8\pi r$, we can write a self-consistent equation for the potential $\phi$ as
\begin{equation}
\phi(r) = \phi_{bare}(r) - \int dr' (gp^i\partial_i\phi(r'))\frac{p^j(r_j-r'_j)}{8\pi|r-r'|},
\end{equation}
where $\phi_{bare}(r) = (p\cdot r)/8\pi r$ is the bare potential associated with the test dipole.  Integrating by parts, we obtain
\begin{align}
\begin{split}
\phi(r) &= \frac{(p\cdot r)}{8\pi r} \\
&- \frac{gp^ip^j}{8\pi}\int dr'\phi(r')\bigg(\frac{\delta_{ij}}{|r-r'|} - \frac{(r_i-r'_i)(r_j-r'_j)}{|r-r'|^3}\bigg).
\end{split}
\end{align}
Taking a Fourier transform, we find
\begin{equation}
\phi(k) = -i\frac{(p\cdot k)}{k^4} - g\phi(k)\frac{(p\cdot k)^2}{k^4}.
\end{equation}
Solving for the potential $\phi(k)$, we have
\begin{equation}
\phi(k) = -i\frac{(p\cdot k)}{k^4+g(p\cdot k)^2} = -i\frac{pk_p}{k_\perp^4 + g p^2k_p^2},
\end{equation}
where $k_p$ is the component of $k^i$ in the $p^i$ direction, and $k_\perp$ is the two-dimensional orthogonal projection.  (We drop the $k_p^4$ term, since it is irrelevant compared to the $k_p^2$ term at small $k$.)  Converting to the interparticle potential energy for dipoles, $V = p^i\partial_i\phi$, we have
\begin{equation}
V(k) = \frac{(pk_p)^2}{k_\perp^4 + gp^2k_p^2}.
\end{equation}
We see that, after screening, $V(k)$ remains bounded at $k=0$, indicating that the long-range interaction has been screened out, leaving only a short-range repulsion between dipoles.


\subsection{Screening of Fractons}

We now investigate the ability of a Fermi surface of dipoles to screen a fracton, which has a much stronger electric field and potential. The potential of a fracton of charge $q$ takes the form $\phi_q = -qr/8\pi$.  The dipoles will still have an induced dipole moment density given by $-p^j(gp^i\partial_i\phi)$, where $\phi$ is the total potential.  Using this information, we can easily modify the self-consistent equation for $\phi$ from the previous section,
\begin{align}
\begin{split}
\phi(r) &= -\frac{qr}{8\pi} \\
&- \frac{gp^ip^j}{8\pi}\int dr'\phi(r')\bigg(\frac{\delta_{ij}}{|r-r'|} - \frac{(r_i-r'_i)(r_j-r'_j)}{|r-r'|^3}\bigg).
\end{split}
\end{align}
After a Fourier transform, we obtain
\begin{equation}
\phi(k) = \frac{q}{k^4} - g\phi(k)\frac{(pk_p)^2}{k^4}.
\end{equation}
Solving for the potential $\phi(k)$, we find
\begin{equation}
\phi(k) = \frac{q}{k_\perp^4 + gp^2k_p^2},
\end{equation}
where we have once again dropped a $k_p^4$ term, which is irrelevant at small $k$.  We now reverse the Fourier transform to find the real-space behavior of the potential,
\begin{equation}
\phi(r) = \int \frac{dk_p d^2k_\perp}{(2\pi)^3} \frac{q}{k_\perp^4 + gp^2k_p^2} e^{i(k_pr_p + k_\perp\cdot r_\perp)}.
\end{equation}
We first set $r_p = 0$, investigating screening transverse to the dipoles.
\begin{align}
\begin{split}
\phi(r_\perp) &= \int \frac{dk_p d^2k_\perp}{(2\pi)^3} \frac{q}{k_\perp^4 + gp^2k_p^2} e^{ik_\perp\cdot r_\perp} \\
&= \frac{q}{2\sqrt{gp^2}}\int \frac{d^2k_\perp}{(2\pi)^2} \frac{1}{k_\perp^2}e^{ik_\perp \cdot r_\perp} \\
&= \frac{q}{4\pi \sqrt{gp^2}}\log r_\perp.
\label{perp}
\end{split}
\end{align}
Similarly, if we set $r_\perp = 0$, studying screening purely along the direction of the dipoles, we obtain
\begin{align}
\begin{split}
\phi(r_p) &= \int \frac{dk_p d^2k_\perp}{(2\pi)^3} \frac{q}{k_\perp^4 + gp^2k_p^2} e^{ik_pr_p} \\
&= \frac{q}{8\sqrt{gp^2}} \int \frac{dk_p}{2\pi} \frac{e^{ik_pr_p}}{k_p}  \\
&= \frac{q}{16\sqrt{gp^2}} \log r_p .
\label{par}
\end{split}
\end{align}
Directions between $r_\perp$ and $r_p$ will behave similarly, with coefficient interpolating between Eqs.~\eqref{perp} and~\eqref{par}.  In either case, we see that the bare linear potential between fractons has been screened down to a logarithmic interaction (with an anisotropic coefficient).  The presence of a dipolar Fermi surface is therefore able to eliminate the issues of ``electrostatic confinement" and allow the fractons to be separated much more easily.


\section{Fermi Statistics of Dipoles}

In the main text, we have often considered the mobile dipoles of fractons to have fermionic statistics.  This is in contrast to some of the simplest lattice models for $U(1)$ fractons~\cite{rasmussen,sub}, in which the dipoles are bosons.  However, there is in principle nothing to prevent the dipoles from being fermions.  Indeed, some of the previously studied discrete fracton theories featured bound states with fermionic statistics~\cite{fracton1}.

One can present field theoretic arguments which indicate that fermionic dipoles should be realizable.  Furthermore, these arguments will teach us how to go about constructing appropriate lattice models.  The primary argument arises from the recently studied ``$\theta$ terms" which can appear in the action for $U(1)$ tensor gauge theories~\cite{chiral}.  These terms have no effect on the gapless gauge mode, but can alter the particle structure of the theory.  A $\theta$ term will attach electric charge to the magnetic monopoles of the theory.  This charge attachment can result in the transmutation of statistics, since electric and magnetic charges pick up phases when moved around each other.  As an example of this phenomenon, in Maxwell theory coupled to bosonic charges, adding an appropriate $\theta$ term will leave the theory almost invariant, with the exception that the magnetic monopole becomes a fermion~\cite{senthil,chong}.

A similar $\theta$ term can be added to the scalar charge theory, which has the effect of attaching dipoles to the magnetic monopoles of the theory (which are vector charges)~\cite{chiral}.  The dipoles and magnetic monopoles pick up phases when moved around each other, so this attachment procedure can produce the same sort of statistical transformations as in conventional Maxwell theory.  By adding an appropriate $\theta$ term, one can convert the dipoles into fermions.  (In the discussion of Ref.~\cite{chiral}, it was always the monopoles which had their statistics changed.  But the electric charge can also have its statistics changed, simply by looking at the problem from the dual perspective.)

At the field theory level, one can conclusively formulate a theory with fermionic dipoles.  As mentioned earlier, there has yet to be a concrete lattice model written down with such particles, but the field theory perspective gives an important clue.  The $\theta$ term in a gauge theory will naturally appear when the charges enter a symmetry protected topological (SPT) phase, protected by time-reversal~\cite{senthil}.  In this specific case, the statistics of the dipoles will be transmuted by placing the magnetic vector charges into a bosonic topological insulator phase.  By tweaking the dynamics of the ``plain-vanilla" lattice models~\cite{rasmussen,sub}, one can drive the monopoles into such a topological insulator state, with the result that the dipoles become fermions.

As an alternative strategy to get fermionic dipoles, one could also consider adapting the strategy of Reference~\cite{ribbon}.  In conventional Maxwell theory, the ground state wavefunction is written in terms of closed string configurations ($i.e.$ a loop gas), and the charges of the gauge field correspond to the endpoints of open strings.  In order to modify the statistics of these charges, one can thicken the strings into ribbons.  By giving the wavefunction an appropriate sign structure, picking up negative signs for each twist of a ribbon, one can thereby change the charges from bosons to fermions~\cite{ribbon}.  A similar strategy will likely work for the case of dipoles.  By adding extra internal structure to the models, along with an appropriate sign structure, one should be able to turn the dipoles into fermions.  Explicitly constructing such a ``ribbon"-like wavefunction is left as a task for future work.


\section{Details on Vector Chern-Simons Theories}

For completeness, here we review some of the established facts regrading Chern-Simons theories of vector gauge fields before calculating the properties of higher rank Chern-Simons gauge theories. 

\subsection{Level Quantization}
\label{quantCS}

In order to establish the quantization of the level $k$ of the Chern-Simons term~\eqref{CS}, we consider the thermal partition function
\beq
\mathcal{Z}[A_\mu] = e^{i \mathcal{S}_{CS}[A_\mu]},
\eeq
by taking time to be Euclidean $\mathbf{S}^1$, parametrized by $\tau \equiv \tau + \beta$, where $\beta$ is the inverse temperature. As mentioned earlier, the action~\eqref{CS} is invariant under $A_\mu \to A_\mu + \p_\mu \alpha$. Let us consider a large gauge transformation which winds around the circle, with $\alpha = 2\pi \tau/\beta$, under which the temporal part of the gauge field gets shifted by a constant,
\beq
\label{gaugetrans}
A_0 \to A_0 + \frac{2 \pi}{\beta}.
\eeq
Furthermore, we now imagine placing the system on a sphere and placing a magnetic monopole inside the sphere (equivalently, threading a background magnetic flux through the sphere), given by
\beq
\int_{\mathbf{S}^2}\,\epsilon^{ij} \p_i A_j = 2\pi,
\eeq
which is the minimum flux allowed by the Dirac quantization condition. Evaluating the Chern-Simons action~\eqref{CS} in such a configuration, and with constant $A_0 = a$, we find that 
\beq
\mathcal{S}_{CS} = \frac{k}{2\pi}\int d^3x\,a \left(\epsilon^{ij} \p_i A_j\right) = ka\beta.
\eeq
Hence, under a gauge transformation of the form~\eqref{gaugetrans}, the action transforms as
\beq
\mathcal{S}_{CS} \to \mathcal{S}_{CS} + 2\pi k,
\eeq
and in order for the partition function $\mathcal{Z}[A_\mu]$ to remain gauge invariant, we require that $k$ is quantized to be an integer, $k \in \mathbb{Z}$.


\subsection{Ground State Degeneracy on a Torus}
\label{sGSD}

Let us consider the Chern-Simons action for the statistical gauge field $a$
\beq
\label{sCS}
\mathcal{S}_{CS} = \frac{m}{4\pi}\int d^3x\, \epsilon^{\mu \nu \lambda} a_u \p_\nu a_\lambda.
\eeq
The equation of motion for the temporal component $a_0$ (or Gauss' Law) is
\beq
\label{gauss}
\epsilon^{ij} \p_i a_j = 0.
\eeq
We now imagine placing this theory on a torus $\mathbf{T}^2 = \mathbf{S}^1\times \mathbf{S}^1$, with the radii of the two circles $R_1$ and $R_2$. On the torus, we can solve the constraint~\eqref{gauss} by setting
\beq
a_i = \frac{\tilde{a}_i}{R_i} + \p_i \Lambda,
\eeq
where $\Lambda$ is periodic on the torus and $\tilde{a}_i$ is spatially constant. Inserting these solutions into the action~\eqref{sCS}, we find that it reduces to 
\beq
\mathcal{S}_{CS} = \frac{m}{4\pi} \int dt\,\epsilon^{ij} \dot{\tilde{a}}_i \tilde{a}_j, 
\eeq
which identifies $\tilde{a}_i$ as the physical degrees of freedom. The canonical commutation relations follow directly from this action,
\beq
[\tilde{a}_1, \tilde{a}_2] = \frac{2\pi i}{m}, 
\eeq
and since the operators $\tilde{a}_i$ are compact, we need to instead consider the algebra of the corresponding gauge-invariant Wilson loop operators,
\beq
W_i = \exp\left(i \oint_{\gamma_i} dx^j \, \tilde{a}_j \right),
\eeq
where $\gamma_i$ are the two non-contractible loops on the torus. The commutation relations of $\tilde{a}$ imply the algebra
\beq
W_1 W_2 = e^{2\pi i/m}W_2 W_1, 
\eeq
the smallest representation of which has dimension $m$,
\beq
W_1\ket{n} = e^{2\pi i n/m} \ket{n},\quad W_2\ket{n} = \ket{n+1}.
\eeq
Thus, the ground state degeneracy of the Chern-Simons action~\eqref{sCS} on a torus is $m$. 


\subsection{Edge Modes of Chern-Simons Theories}
\label{sEdge}

Let us consider the Chern-Simons action~\eqref{sEFT} which describes the FQHE state at filling fraction $\nu = 1/m$,
\beq
\mathcal{S}_{CS} = \frac{m}{4\pi} \int d^3x\, \epsilon^{\mu \nu \lambda} a_\mu \p_\nu \lambda.
\eeq
In order to study the edge excitations of this system, we imagine placing it on a semi-infinite plane with a boundary at $y=0$ such that the quantum Hall fluid lives at $y<0$ with vacuum at $y>0$. Under a gauge transformation $a_\mu \to a_\mu + \p_\mu \alpha$, the action transforms as
\beq
\mathcal{S}_{CS} \to \mathcal{S}_{CS} + \frac{m}{4\pi} \int_{y = 0} dxdt\, \alpha \left(\p_t a_x -  \p_x a_t \right),
\eeq
i.e., it is gauge-invariant only up to a surface term. Thus, we require additional degrees of freedom living at the edge in order to have a fully gauge-invariant theory.

In order to deduce these additional degrees of freedom, we consider the variation of the action~\eqref{sEFT} in the presence of a boundary,
\beq
\delta \mathcal{S}_{CS} = \frac{m}{4\pi} \int d^3x\, \epsilon^{\mu \nu \lambda} \left(\delta a_\mu f_{\nu \lambda} + \p_\mu \left(a_\nu \delta a_\lambda \right) \right),
\eeq
where $f_{\mu \nu} = \p_\mu a_\nu - \p_\nu a_\mu$. Thus, we see that minimizing the action leads to the required equation of motion $f_{\nu \lambda} = 0$ (equivalently, the zero flux condition) only if we can set the second term to zero. We can achieve this by setting $a_t = 0$ at the boundary $y=0$.

Now, in order to derive the action for the boundary degrees of freedom, we extend this boundary condition into the bulk i.e., we fix the gauge $a_t = 0$ in the bulk. Then, the constraint imposed by the gauge-fixing condition remains
\beq
\epsilon^{ij} \p_i a_j = 0,
\eeq
which is solved in terms of a compact scalar field $\varphi$ by taking $a_i = \p_i \varphi$. Inserting this into the action $\mathcal{S}_{CS}$, with $a_t = 0$, we obtain the edge action
\beq
\mathcal{S}_{CS} = \int_{y=0} dxdt\, \p_t \varphi \p_x \varphi.
\eeq

This is, however, not the most general action we could write down at the edge since we can add energetic terms which respect the shift symmetry $\varphi \to \varphi +$ constant. A mass term $\sim \varphi^2$ term is prohibited (even in the absence of U(1) symmetry) since the field $\varphi$ is non-local, as evidenced by the commutation relation
\beq
[\varphi(x),\varphi(x')] = \frac{i\pi}{m} \text{sgn}(x - x').
\eeq
Including the lowest order spatial derivatives, we find the general edge action for a chiral Chern-Simons theory
\beq
\mathcal{S}_{\text{edge}} =  \int_{y=0} dxdt\, \left(\p_t \varphi \p_x \varphi - v \p_x \varphi \p_x \varphi\right),
\eeq
where $v$ is a non-universal velocity which depends on the microscopic details of the edge. The equation of motion satisfied by $\varphi$ is
\beq
\p_t \p_x \varphi - v \p_x \p_x \varphi = 0.
\eeq 
Re-stated in terms of the field $\rho = 1/(2\pi) \p_x \varphi$,
\beq
(\p_t  - v \p_x) \rho = 0,
\eeq
we find the equation governing a chiral density wave propagating along the boundary with velocity $v$. By coupling the action $\mathcal{S}_{\text{CS}}$, one can show that $\rho$ is indeed the charge density along the edge and that the operator describing electrons at the boundary is
\beq
\Psi = :e^{im\varphi}:
\eeq
where the colons denote normal ordering. We refer the reader to the excellent reviews~\cite{wenbook,tong} for a more comprehensive discussion of these boundary CFTs. 


\section{Details on Generalized Chern-Simons Theories}
\label{gendet}

Here, we derive some of the technical details regarding higher rank U(1) Chern-Simons theories described by a traceless symmetric rank 2 tensor. The action for such a theory is 
\beq
\begin{split}
\mathcal{S}_{gCS}[\tilde{A}_{ij},\phi] =& -\frac{\theta}{4\pi^2} \int d^3x\, \phi \epsilon^{bi} \p_i \p_j \tilde{A}_b^{\,\,\,j} \\
 &+ \frac{\theta}{8\pi^2}\int d^3x\, \epsilon^{bi} \dot{\tilde{A}}_{ij} \tilde{A}_b^{\,\,\,j}.
\end{split}
\eeq
A remark regarding the dimensions of these fields: in units where $\hbar = c = 1$, $\tilde{A}_{ij}$ has the same units as those of a vector gauge field $A_i$,
\beq
[\tilde{A}_{ij}] = [A_i] = L^{-1},
\eeq
where $L$ denotes length. However, while the temporal component of a vector gauge field also has dimensions $L^{-1}$, the Lagrange multiplier field $\phi$ has dimension
\beq
[\phi] = L^{0}.
\eeq
Since under a gauge transformation $\phi$ transforms as $\phi \to \phi + \dot{\alpha}$, this implies that the gauge parameter $\alpha$ has dimensions of length,
\beq
[\alpha] = L^{1},
\eeq
consistent with the gauge transformation and dimension of $\tilde{A}_{ij}$.

\subsection{Quantization of $\theta$}
\label{quant}
Recall that the magnetic flux vector $B^j = \epsilon^{ib} \p_i \tilde{A}_b^{\,\,\,j}$ in the higher rank theory~\eqref{geneft} is one-dimensional, moving only transversely i.e., on a square lattice, $B^x$ may only move in the $y$-direction. If we now place the system on a sphere, we can imagine threading a flux through this sphere. The minimum such flux allowed by the Dirac quantization condition is
\beq
\label{flux}
\int_{\mathbb{S}^2} B^j = 2\pi \hat{x}^j, 
\eeq
where $\hat{x}^j$ simply reflects the vector nature of the magnetic flux in this theory.

Generalizing the discussion of level quantization in Appendix~\ref{quantCS}, we consider the thermal partition function 
\beq
Z_{gCS}[\tilde{A}_{ij},\phi] = e^{i S_{gCS}[\tilde{A}_{ij},\phi]}
\eeq
by taking time to be Euclidean $\mathbf{S}^1$, parametrized by $\tau \equiv \tau + \beta$, where $\beta$ is the inverse temperature. As established earlier, the theory~\eqref{geneft} is invariant under the gauge transformation $\phi \to \phi + \dot{\alpha}$. Since the field $\phi$ acts as a Lagrange multiplier and has no dynamics of its own, its role in the theory is analogous to that of $A_0$, the temporal component of a vector gauge field. However, as mentioned earlier, $\phi$ is dimensionless and we should instead consider a transformation of $\p_i \phi$, which has the same dimensions as $A_0$: $[\p_i \phi] = L^{-1}$. We hence consider a large gauge transformation
\beq
\alpha(\vec{x},t) = \frac{2 \pi \tau}{\beta}(x_1 + x_2),
\eeq
under which $\p_i \phi$ transforms as
\beq
\label{gtrans}
\p_i \phi \to \p_i \phi + \frac{2 \pi}{\beta} \hat{x}_i.
\eeq
Evaluating the generalized Chern-Simons action~\eqref{geneft} on a configuration specified by the flux threading condition~\eqref{flux} and with constant $\p_i \phi = \Phi \hat{x}_i$, we find 
\beq
S_{gCS} = -\frac{\theta}{4\pi^2}\Phi \hat{x}_j \int B^j = -\frac{\theta}{2\pi}\Phi \beta.
\eeq 
Under the large gauge transformation specified by Eq.~\eqref{gtrans}, the action shifts by a constant
\beq
S_{gCS} \to S_{gCS} - \theta,
\eeq
and hence, in order for the partition function to be gauge-invariant, the coefficient $\theta$ must be quantized in units of $2\pi$, 
\beq
\theta = 2 \pi k,\quad k \in \mathbb{Z}.
\eeq

\subsection{Ground State Degeneracy}
\label{GSD}

The generalized Chern-Simons action for an emergent FQH state of dipoles at filling fraction $\nu = 1/m$ is
\beq
\label{appeft}
\mathcal{S}_{gCS} = -\frac{m}{2\pi} \int d^3x\, \chi\,\epsilon^{bi} \p_i \p_j a_b^{\,\,\,j} + \frac{m}{4\pi}\int d^3x\,\epsilon^{ib}\dot{a}_{ij}a_b^{\,\,\,j},
\eeq 
where $a_{ij}$ is a traceless symmetric tensor. On a closed manifold, this theory is gauge-invariant under
\beq
\begin{split}
a_{ij} &\to a_{ij} + \left(\p_i \p_j - \frac{1}{2}\delta_{ij} \delta^2 \right)\alpha, \\
\chi & \to \chi + \dot{\alpha},
\end{split}
\eeq
for arbitrary gauge parameter $\alpha(\vec{x},t)$. The Gauss' law constraint for this theory, enforced by the Lagrange multiplies field $\chi$, is
\beq
\label{gausslaw}
-\frac{m}{2\pi} \epsilon^{bi} \p_i \p_j a_b^{\,\,\,j} = 0.
\eeq
We note that the two independent components of $a_{ij}$ are combined into one canonically conjugate pair, with commutation relations
\beq
[a_{xx}(\vec{x}),a_{xy}(\vec{x}')] = -\frac{i \pi}{m} \delta(\vec{x} -\vec{x}').
\eeq
Due to the Gauss' law constraint on $a_{ij}$, the theory is fully constrained and there are no local degrees of freedom.

We now imagine placing this system on a torus $\mathbf{T}^2 = \mathbf{S}^1 \times \mathbf{S}^1$, with the radii of the two circles $R_1$ and $R_2$. 
On an $R_1\times R_2$ torus, we can satisfy the Gauss' law constraint~\eqref{gausslaw} by setting the two conjugate variables to 
\beq
\begin{split}
a_{xx} = -a_{yy} &= \frac{a(t)}{R_1 R_2} + \frac{1}{2}(\p_x^2 - \p_y^2)\Lambda,\\
a_{xy} = a_{yx}  &= \tilde{a}(t) + \p_x\p_y \Lambda,
\end{split}
\eeq
where $\Lambda(\vec{x},t)$ is an arbitrary continuous, periodic function on the torus. Note that $a$ and $\tilde{a}$ only describe the topological contribution to the action, since we have separated them from the gauge-redundant part $\Lambda$. Inserting these solutions into the action~\eqref{appeft}, we find
\beq
\mathcal{S}_{gCS} =\frac{m}{\pi} \int dt\,\dot{a}(t)\tilde{a}(t) .
\eeq
This theory has a ground state degeneracy~\cite{wenzee2,slagle2}
\beq
\text{GSD} = 2m.
\eeq


\newpage

\bibliography{library}



\end{document}